\documentclass[a4paper,10pt]{article}
\pdfoutput=1 

\usepackage{jheppub} 

\usepackage[T1]{fontenc} 

\usepackage{physics}
\usepackage{dsfont}
\usepackage[normalem]{ulem}

\usepackage[title,titletoc]{appendix}

\newcommand{\mn}[1]{}

\title{A $T \bar{T}$ Deformation for Curved Spacetimes from 3d Gravity}

\author[a]{Edward A. Mazenc,}
\author[b]{Vasudev Shyam}
\author[a]{and Ronak M. Soni}
\affiliation[a]{Stanford Institute for Theoretical Physics, 382 Via Pueblo, Stanford CA 94305 }
\affiliation[a]{Perimeter Institute for Theoretical Physics,
31 Caroline St. N, N2L 2Y5, Waterloo ON, Canada}

\vspace{5mm}

\vspace{1cm}

\emailAdd{mazenc@stanford.edu}
\emailAdd{vshyam@perimeterinstitute.ca}
\emailAdd{ronakms@stanford.edu}

\abstract{ 
We propose a generalisation of the $T \bar{T}$ deformation to curved spaces by defining, and solving, a suitable flow equation for the partition function. We provide evidence it is well-defined at the quantum level.
This proposal identifies, for any CFT, the $T \bar{T}$ deformed partition function and a certain wavefunction of 3d quantum gravity.
This connection, true for any $c$, is not a holographic duality --- the 3d theory is a ``fake bulk.''
We however emphasise that this reduces to the known holographic connection in the classical limit.

Concretely, this means the deformed partition function solves exactly not just one global equation, defining the $T \bar{T}$ flow, but in fact a local Wheeler-de Witt equation, relating the $T \bar{T}$ operator to the trace of the stress tensor. This also immediately suggests a version of the $T \bar{T}$ deformation with locally varying deformation parameter.

We flesh out the connection to 3d gravity, showing that the partition function of the deformed theory is precisely a 3d gravity path integral.
In particular, in the classical limit, this path integral reproduces the holographic picture of Dirichlet boundary conditions at a finite radius and mixed boundary conditions at the asymptotic boundary.

Further, we reproduce known results in the flat space limit, as well as the large $c$ $S^2$ partition function, and conjecture an answer for the finite $c$ $S^2$ partition function.

}

\begin{document}
\maketitle
\flushbottom

\section{Introduction} \label{intro}
In flat space, the $T \bar{T}$ deformation has been discovered at least five times  \cite{Zamolodchikov:2004ce,Smirnov:2016lqw,Lechner:2006kb,Dubovsky:2012wk,Freidel:2008sh}. This serves as a testimony to the many deep facts we may hope to learn from a better understanding of the deformation and its generalization to curved space. So far, each incarnation has revealed novel features. \cite{Zamolodchikov:2004ce, Smirnov:2016lqw} highlighted its solvability and the way it preserves the seed theory's integrability. $T \bar{T}$ being an irrelevant operator, the flow generated by the deformation provides an exceptional example of what resembles flowing up an renormalization group (RG) trajectory \cite{ZamTalk}.\footnote{Though, as emphasized in \cite{Komargodski:talk}, it is not precisely an RG flow, since we vary only the $T\bar{T}$ coupling but keep all other couplings in the theory fixed.} Analysing modular transformation properties of the partition function, \cite{Datta:2018thy,Aharony:2018bad,Aharony:2018ics} pointed out how the $T\bar{T}$ and related $J \bar{T}$ deformation were singled out by very few  assumptions. Further, the deformed theory's energy spectrum indicates it cannot be reduced to a local quantum field theory.\footnote{For one sign of the deformation, the spectrum is Hagedorn. For the other, infinitely many energy levels become complex. \cite{Cardy:2019qao,Lewkowycz:2019xse} also made precise statements regarding its non-locality at the level of operator algebras. }   \cite{Cardy:2018sdv,Dubovsky:2017cnj,Dubovsky:2018bmo} showed the deformed theory on a torus was in fact a theory of two-dimensional quantum gravity. Further evidence the $T\bar{T}$ theory cannot be a simply local field theory came from the connection to string theory exhibited in \cite{Dubovsky:2017cnj,Callebaut:2019omt}. First indications of a higher-dimensional interpretation of the deformed 2d theory rested upon a connection to holographic RG \cite{McGough:2016lol,Shyam:2017znq,Kraus:2018xrn,Hartman:2018tkw}. Higher-dimensional generalisations of the deformation itself, i.e. deformations of field theories living in $d>2$, proceeded by replicating this property \cite{Hartman:2018tkw,Taylor:2018xcy}. It is also possible to define lower dimensional deformations, one of which reproduces this connection \cite{Gross:2019ach,Gross:2019uxi}.
Finally, it should also be noted that closely related deformations can be more directly related to string theory \cite{Giveon:2017nie,Giveon:2017myj}.
A helpful introduction with a more complete set of references can be found in \cite{Jiang:2019hxb}.

The usual definition of the $T \bar{T}$ deformation has been logically predicated upon the result of \cite{Zamolodchikov:2004ce}. It explains how this bilinear operator is naturally well-defined in any theory with a local conserved stress tensor, global translation-invariance and with a least at one non-compact spacetime direction.\footnote{\cite{Zamolodchikov:2004ce} in fact requires one additional, rather technical condition, see his section 4.} 
Once assured the operator exists, we can use it to define a deformation equation,
\begin{equation}
  \partial_{\lambda} \log Z_{\lambda} = \int \sqrt{g} \langle T \bar{T}_{\lambda} \rangle_{\lambda}, \quad T \bar{T}_{\lambda} = \det T_{\lambda} = \frac{1}{2} \epsilon^{\mu\nu} \epsilon^{\rho\sigma} T^{\lambda}_{\mu\rho} T^{\lambda}_{\nu\sigma}.
  \label{eqn:ttbar-defn}
\end{equation}
The purpose of the two $\lambda$ subscripts is to indicate that it is the expectation value of the stress tensor of the $\lambda$-deformed theory \emph{in} the vacuum of the $\lambda$-deformed theory.

The objective of this paper is to propose a quantum-mechanically well-defined generalization of the $T\bar{T}$ deformation for two-dimensional seed theories defined on curved spaces.
The reason this is a conceptually distinct problem from the flat-space case is that we no longer have an analogue of the theorem of \cite{Zamolodchikov:2004ce,Smirnov:2016lqw}. Indeed, a crucial ingredient there was a certain translation-invariance assumption that has no hope of being true on curved manifolds. In fact, direct investigation suggests that no simple analogue of their theorem can hold in curved space \cite{Jiang:2019tcq}.
Given this fact, the approach of this paper is somewhat backwards. We first posit the form of the deformed theory's partition function, then show it satisfies a particular differential equation (and initial condition) we deem a reasonable curved-space generalisation of the $T \bar{T}$ flow. 

A particular object in 3d gravity --- known as a radial wavefunction --- provides the expression for the deformed theory's partition function, which we denote by $Z_{\lambda}[f]$ for reasons to be explained momentarily. This connection relies crucially on the work of \cite{Freidel:2008sh}, who showed how a certain integral kernel mapped CFT partition functions to these radial wavefunctions.
These radial wavefunctions satisfy an infinite set of \emph{local} constraints known as the radial Hamiltonian constraint, also known as the Wheeler-de-Witt equation.
As such, $Z_{\lambda}[f]$ unexpectedly satisfies not just the global flow equation fixing its dependence on $\lambda$ but an infinite set of equations.
These equations relate the expectation value of the $T\bar{T}$ operator at a point to the one point function of the trace of the stress tensor at the same point. They therefore provide us with partial control on the divergence structure of the $T \bar{T}$ operator.
We would like to take another sentence to emphasise that we do not know anything general about any reasonable definition of the $T\bar{T}$ operator in other theories, including the QFTs that provide the seeds for the flow.

\begin{figure}[h]
    \centering
    \includegraphics[width=50mm]{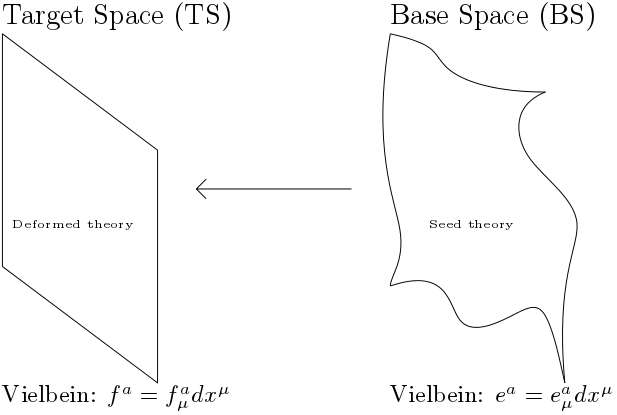}
    \caption{The main set-up of the deformation. The seed QFT lives on a dynamical base space, whose metric is coupled to the fixed target space. The target space is the space on which the deformed theory lives.}
    \label{fig:ts-bs}
\end{figure}

The main claim we make is that for any seed QFT with partition function $Z_0$, we can write the deformed theory's partition function via\footnote{We direct the reader unfamiliar with vielbeins to appendix \ref{app:e} for a short introduction.}
\begin{equation}
    Z_{\lambda}[f] = \int De \ e^{-\frac{1}{2 \lambda} \int \varepsilon_{ab} (f-e)^a \wedge (f-e)^b} Z_0[e],
    \label{eqn:kernel-intro}
\end{equation}
where $Z_0[e]$ stands for the partition function of the undeformed theory and $\lambda$ is the deformation parameter. Crucially, the seed theory lives on a 2d base space with metric parameterized in terms of vielbeins $e^a_{\mu}$, while the deformed theory resides on a target space with metric $ds^2=\delta_{ab}f^a_{\mu}f^b_{\nu}dx^{\mu}dx^{\nu}$, see figure \ref{fig:ts-bs}.  Beyond requiring the base and target share the same topology, we have yet to find further restrictions on the vielbeins $f$. 
This partition function satisfies the flow equation
\begin{equation}
    \partial_\lambda Z_\lambda [f] = \frac{1}{2} \int d^2 x  :\varepsilon_{\mu\nu} \varepsilon^{ab} \frac{\delta}{\delta f_\mu^a (\sigma)} \frac{\delta}{\delta f_\nu^b (\sigma)}: Z_\lambda [f],
    \label{eqn:flow-intro}
\end{equation}
where we define the coincident double derivative not as a limit but merely by the subtraction of a contact term.
It is not obvious that this flow equation, with this definition of the deformation operator, is \emph{the} curved space generalisation of the $T \bar{T}$ deformation. A more principled approach might indeed give rise to a theory under better analytic control and with more desirable properties.

More importantly, it is not even completely obvious so far that our proposal is even \emph{a} sensible generalisation.
To wit, there's no guarantee that the path integral \eqref{eqn:kernel-intro} can be suitably regularised as a functional of the metric.
While the precise consistency conditions are not clear at the moment, a minimal one is that the partition function of a compact Euclidean manifold is rendered finite by a single UV cutoff and a finite number of counterterms.
This is equivalent to saying $n$-point functions of the stress tensor have singularities that can be integrated against metric perturbations.
While we will not be able to prove that this is the case for our $Z_{\lambda}[f]$, we will provide some evidence for it.

 
One of the major pieces of evidence for this comes from the connection to 3D gravity. This extends to the fully quantum setting that which was already understood by \cite{McGough:2016lol,Shyam:2017znq,Kraus:2018xrn,Hartman:2018tkw,gorbenko2019ds}.
This connection is rooted in a relation between the two-dimensional ``trace flow equation'' and one of the equations of motion of three-dimensional general relativity.
Consider 3d Einstein gravity with a negative cosmological constant. Imagine the manifold is sliced with a radial coordinate $\rho$ and 2d slices transverse to it that are parameterized by $x^\mu$.
In terms of the Brown-York stress tensor,\footnote{We have added a Balasubramanian-Kraus counterterm \cite{Balasubramanian:1999re}, and will keep it around for most of this paper.}
\begin{equation}
    T_{\mu\nu} = \frac{2}{\sqrt{|g_{(2)}|}} \frac{\delta S_{GR}}{\delta g_{(2)}^{\mu\nu}} = \frac{1}{8\pi G_N} \left\{ K_{\mu\nu} - \left(K - \frac{1}{l}\right) g_{\mu\nu} \right\}
    \label{eqn:by-T}
\end{equation}
the $\rho,\rho$ component of the Einstein equations can be rewritten as
\begin{equation}
    T^\mu_\mu = 8\pi G_N l \det T - \frac{l}{16\pi G_N} R_{(2)}.
    \label{eqn:e-rho-rho}
\end{equation}

A very similar equation can be obtained by making an assumption about the $T \bar{T}$-deformed theory based on dimensional analysis.
For simplicity, let us deform a seed CFT by $\delta S = \lambda \int d^2x T \bar{T}$. This deformation parameter $\lambda$ is the only mass scale in the theory. We might thus assume the response of the partition function to an overall scaling is given entirely by its response to a change in $\lambda$ alongside the contribution from the conformal anomaly,
\begin{equation}
  \int \langle T^{\mu}_{\mu} \rangle_{\lambda} - \lambda \partial_{\lambda} \log Z_{\lambda} = \frac{c}{48\pi} \int R.
  \label{eqn:dim-anal-eqn}
\end{equation}
Plugging the deformation equation into this equation, we find
\begin{equation}
  \int \langle T^{\mu}_{\mu} \rangle_{\lambda} = \lambda \int \langle T \bar{T} \rangle_{\lambda} + \frac{c}{48\pi} \int R.
  \label{eqn:wdw-eqn}
\end{equation}
This is tantalisingly similar to \eqref{eqn:e-rho-rho}, but not quite the same.

Let us list out the differences.
First, the coupling constants are different. This can easily be dealt with by positing a dictionary relating $(c,\lambda)$ to $(G_N,l)$.
Secondly, \eqref{eqn:e-rho-rho} is valid point-wise, while \eqref{eqn:wdw-eqn} resides under an integral sign. 
Third, \eqref{eqn:e-rho-rho} is a classical relation, but \eqref{eqn:wdw-eqn} is a quantum one; we posit the latter as a quantisation of the first, classical relation. More precisely, quantising a theory of gravity requires the imposition of certain equations of motion as constraints identically satisfied by all physical wavefunctions; these constraints are known as the Hamiltonian constraint or the Wheeler-de Witt (WdW) equation,
\begin{equation}
  H_{WdW} \Psi [g^{(2)} (\rho)] = 0, \quad H_{WdW} = \hat{E}_{\rho}^{\rho}.
  \label{eqn:radial-h}
\end{equation}
It is worth noting the unconventional feature that the ``wavefunction'' that appears in this equation is a state on a 2d \emph{spacetime} with the radial direction playing the role of time --- this object is known as a radial wavefunction.

The important point about this connection with the WdW equation is that the heuristic derivation works for any CFT, independent of its central charge or field content.
While, for a holographic CFT in the stress tensor sector, this deformation appears to define the theory with Dirichlet boundary conditions at finite radius in the dual $AdS$, the point is that \emph{for every CFT} the deformation behaves like flowing into a quantum ``fake bulk.''\footnote{This point was first explained to us by Aitor Lewkowycz, and it will be expanded upon in \cite{Belin:unp}.}
This ``fake bulk'' is not a quantum holographic dual, but it does agree with holography in the classical limit.

As it happens, a radial wavefunction satisfying \eqref{eqn:dim-anal-eqn} -- \eqref{eqn:radial-h} was already found in \cite{Freidel:2008sh}. Minor rescalings lead immediately to \eqref{eqn:kernel-intro}.
Briefly, \cite{Freidel:2008sh} proved that, given a 2D CFT partition function $Z_{0}[e]$ as a function of the 2D vielbein $e_{\mu}^{a}$ that satisfies the conformal anomaly equation
\begin{equation}
  e_{\mu}^{a} \frac{\delta}{\delta e_{\mu}^{a}} \log Z_{0} [e] = \frac{c}{24\pi} \det(e) R_{e},
  \label{eqn:anomaly-eqn}
\end{equation}
\eqref{eqn:kernel-intro} satisfies the radial WdW equation \eqref{eqn:radial-h}.
This result is exact, and does not depend on any large $c$ limit.

The central point of this paper is that this same object that satisfies the Wheeler-de Witt (WdW) equation also satisfies the flow equation.
Whereas it satisfies the WdW equation only when $Z_0$ is a CFT partition function satisfying \eqref{eqn:anomaly-eqn}, it satisfies the flow equation for any $Z_0$.
In other words, even though the connection to 3d GR is only true when the seed $Z_0$ is a CFT partition function, the flow equation is indifferent to the nature of $Z_0$.
Thus, the theory \eqref{eqn:kernel-intro} is a proposal for the curved space deformed theory for any QFT.

We present the deformed theory first from a purely 2d perspective by exhibiting  various properties the kernel satisfies. In particular, we show
\begin{enumerate}
    \item the deformed theory reduced to the undeformed one in the limit $\lambda \to 0$.
    \item Its stress tensor is conserved and symmetric, which expresses the invariance of $Z_{\lambda}[f]$ under tangent space rotations and diffeomorphisms of the manifold equipped with the vielbeins $f$.
    \item The integral kernel composes. Plugging in a theory already deformed by $\lambda'$ as the seed results in a theory deformed by $\lambda + \lambda'$.
    \item At leading order in large $c$, where the kernel has a classical limit, it reproduces previously known $S^2$ partition functions and entanglement entropies.
    \item The loop expansion is controlled by a renormalisable coupling $1/\lambda$, suggesting that the theory can be regulated with a finite number of counterterms.
\end{enumerate}

Further, our analysis immediately suggests an important further generalisation of the deformation. The standard flow equation is only related to the integral of the WdW equation, not its local form. Since $Z_{\lambda}[f]$ solves the local WdW  equation, we may in fact easily modify the kernel to promote $\lambda \rightarrow \lambda(x)$. 
This local deformation also satisfies a different local WdW equation.

We then move on to flesh out the connection with 3d gravity, showing precisely how the kernel can be thought of as the path integral on an annular region with a Dirichlet boundary condition on one side and a CFT-dependent boundary condition on the other. We also show how many of the properties of the 2d theory are natural from this point of view.

We also, using a particular gauge-fixing of the 3d gravity wavefunction, conjecture --- up to ignorance parameters --- an exact answer for the finite $c$ deformed partition function on an $S^2$.
Most interestingly, for a particular value of the ignorance parameter, it can be written as a localised version of the Freidel kernel.

Finally, we discuss some objections to the existence of the curved space deformation. We outline several arguments that it should not exist and highlight to what extent we have shown that the kernel evades those issues. A by-product of this discussion is the observation that a quantum version of the $T \bar{T} + OO$ deformations of \cite{Hartman:2018tkw,Taylor:2018xcy} may in fact exist as well.

The plan of this paper is as follows.
In section \ref{2d}, we carefully define the theory as a partition function, show some basic properties, and explain the natural way to obtain other deformations.
Then, in section \ref{3dgravity}, we flesh out the picture of the kernel as a radial wavefunction and recast the above properties of the 2d theory in 3d language.
We proceed to obtain the CMC gauge-fixed genus sphere partition function --- which we conjecture to be the exact $S^2$ partition function for a CFT --- in section \ref{sec:cmc}.
We also comment on the large $c$ limit of the un-gauge-fixed $S^2$ partition function in section \ref{sec:s2}.
Finally, we address the expectation that the curved space theory should not exist and the extent to which we have addressed these concerns in section \ref{sec:arg}.
We conclude in section \ref{sec:conc}, sketching future directions of work.

{\bf{Note Added:}}
While this manuscript was in preparation, the preprint \cite{Tolley:2019nmm} with some overlapping results appeared on the arXiv.

\subsection*{A Note on Notation}
Firstly, we will liberally use the first-order formalism in this paper, since that is the one in which the kernel is simplest. For those unfamiliar with this formalism, we have included a short introduction to vielbeins in appendix \ref{app:e}.

Secondly, we have found ourselves in the unfortunate situation of having to deal with both Levi-Civita tensor densities as well as Levi-Civita symbols.
We will consistently use the notation $\epsilon$ for the tensor density and $\varepsilon$ for the symbol,
\begin{align}
    \epsilon_{01} &= \sqrt{g} \nonumber\\
    \varepsilon_{01} &= 1. 
    \label{eqn:eps-not}
\end{align}
We will explicitly state which one turns up in an equation when the difference is important.

\subsection*{A Note on A Sign}
Throughout this paper, we work with $\lambda > 0$ being the holographic sign of the deformation.
While this is widely considered the bad sign, we find it useful for two reasons.
The first is that much of the story in this paper relates to 3d gravity, which is somewhat more confusing for the other sign.
The second is that this is the sign for which the kernel \eqref{eqn:kernel-intro} does not obviously have a conformal mode problem.

\section{The 2D Story: Flows of Partition Functions} \label{2d}
The $T\bar{T}$ deformation defines a flow for the partition function. Analysing this flow via an integral kernel, i.e. rewriting the deformed theory partition function as a path integral transform of the seed's, dates back to \cite{Cardy:2018jho}. It has since appeared in multiple guises, amongst others in the works of \cite{Dubovsky:2017cnj,Dubovsky:2018bmo,McGough:2016lol,Hashimoto:2019wct,Tolley:2019nmm,Ireland:2019vvj}. 
In this section, we present the deformation purely in terms of an integral kernel and show what properties it satisfies. It conveniently circumvents several of the issues raised in generalizing the precise $T\bar{T}$ operator to curved space. While many of the kernel's remarkable properties are most easily seen from a 3d gravity perspective, we first present our results in their immediate 2d setting. 

\subsection{A closer look at the Kernel} \mn{this stuff is important, but technical. someone reading this stuff for first time probably  really care. Maybe this should go into an appendix instead? Was just trying to make it clear we had thought carefully about this path integral and the subtleties we face in arguing for the well-posedness of the whole thing}
To argue for the well-definedness of the proposed deformation, the path integral transform quoted in the introduction
\begin{equation}
  Z_{\lambda} [f] = \int De\ e^{- \frac{1}{\lambda} \int (f-e)^{1} \wedge (f-e)^{2}} Z_{0} [e].
  \label{eqn:laurent-kernel}
\end{equation}
requires a careful specification of the measure $De$ and seed partition function $Z_0[e]$.

The measure for the integration over vielbeins, elaborated upon in appendix \ref{app:De}, is both diffeomorphism- as well as translation-invariant,\footnote{We thank A. Tolley for pointing out the existence of this measure.}
\begin{equation}
    De = De^\xi = D(e+c).
    \label{eqn:De-props}
\end{equation}
The existence of this measure is a special fact about 2d gravity in the first-order formalism.

This measure ensures that the seed partition function reappears from the kernel in the limit $\lambda \rightarrow 0$. This choice is also required for the kernel to compose, that is, deforming by $\lambda_1$ and again by $\lambda_2$ is identical to doing a single deformation by $\lambda_1 + \lambda_2$. 

We take $Z_0[e]$ to be defined via its path integral formulation. Even for a CFT, this involves a particular regularization scheme, such as the choice of a local-counter term  canceling off any contribution to the cosmological constant (see for example \cite{polchinski1986evaluation}). This matches the implicit prescription of \cite{Dubovsky:2018bmo} for the case of the torus. We will further assume it is invariant under background spacetime diffeomorphism $Z_{0}[e]=Z_{0}[e^{\xi}]$, where $\xi$ parameterizes the diffeomorphism.

\subsection{The Flow Equation}
In this subsection, we show that the Freidel kernel \eqref{eqn:laurent-kernel} satisfies a particular generalisation of the $T \bar{T}$ deformation equation \eqref{eqn:ttbar-defn}, independent of the nature of the seed.
The most important point is that we do not have an a priori definition of the $T \bar{T}$ operator that appears in this equation, but simply that the kernel satisfies it.

The flow equation satisfied by the kernel is
\begin{equation}
    \partial_{\lambda} Z_{\lambda} [f] = \int d^2 x \left( \frac{1}{2} \varepsilon^{ab} \varepsilon_{\mu\nu} :\frac{\delta}{\delta f_{\mu}^{a}(x)} \frac{\delta}{\delta f_{\nu}^{b}(x)}: \right) Z_{\lambda} [f],
    \label{eqn:flow-eqn}
\end{equation}
where the `normal ordering' is defined not by a coincident limit of any sort but simply as
\begin{equation}
    \frac{1}{2} \varepsilon^{ab} \varepsilon_{\mu\nu} :\delta_{f_\mu^a (x)} \delta_{f_\nu^a (x)}: = \frac{1}{2} \varepsilon^{ab} \varepsilon_{\mu\nu} \delta_{f_\mu^a (x)} \delta_{f_\nu^a (x)} + \frac{2}{\lambda} \delta^{(2)}(0). 
    \label{eqn:normal-ordering}
\end{equation}
This is all the normal ordering one needs to do to get the flow equation \eqref{eqn:flow-eqn} to work; if it turns out that there are more divergences on the RHS, they also drive this flow.

Before proving the equation, let us define the one- and two-point functions of the stress tensor.
The one-point function is defined by\footnote{Despite the factor of $2$, this is consistent with \eqref{eqn:by-T}. The factor is absorbed by the transformation between metrics and vielbeins.}
\begin{equation}
  \langle T^{\mu}_{a}(x) \rangle = - (\det f(x))^{-1} \frac{1}{Z[f]} \frac{\delta}{\delta f_{\mu}^{a}(x)} Z[f],
  \label{eqn:T-defn}
\end{equation}
which means that
\begin{equation}
    \delta (-\log Z) = \int (\det f) \delta f_\mu^a \langle T^\mu_a \rangle.
    \label{eqn:T-defn-logic}
\end{equation}
Similarly, the two-point function is defined as
\begin{equation}
    \langle T^\mu_a(x) T^\nu_b(y) \rangle = \frac{1}{Z[f]} \frac{1}{\det f(x) \det f(y)} \delta_{f_\mu^a (x)} \delta_{f_\nu^b (y)} Z[f],
    \label{eqn:TTdefn}
\end{equation}
where the $\det f$ factors are outside so that the change in the free energy is a double integral of the two-point function.
With this definition, the RHS of \eqref{eqn:flow-eqn} (up to the normal ordering) is \footnote{Recall that $\epsilon$ is a tensor density. $\varepsilon$ on the other hand is simply the Levi-Civita symbol. Hence, $\epsilon_{\mu \nu} = \det(f) \varepsilon_{\mu \nu}$.}
\begin{equation}
    \int d^2 x (\det f(x))\ \varepsilon^{ab} \epsilon_{\mu\nu} \frac{1}{(\det f(x))^2} \delta_{f_\mu^a} \delta_{f_\nu^b} Z_\lambda = Z_\lambda \int d^2 x \det f \varepsilon^{ab} \epsilon_{\mu\nu} \langle T^\mu_a T^\nu_b \rangle
\end{equation}
Thus, it appears a sensible generalisation of the $T \bar{T}$ operator.

Moving on to prove the kernel satisfies the flow equation, we see the left hand side of \eqref{eqn:flow-eqn} becomes simply
\begin{align}
  \partial_{\lambda} Z_{\lambda}[f] = \int De  \left( \frac{1}{2 \lambda^2} \int \varepsilon_{ab} (f-a)^{a} \wedge (f-e)^{b} \right) e^{- S_{K,\lambda}} Z_{0} [e].
  \label{eqn:flow-lhs}
\end{align}

Let us first work out the right hand side, without the normal-ordering, keeping any new contact terms which may arise:
\begin{align}
  \frac{1}{2} \int d^{2} x\ \varepsilon^{ab} \varepsilon_{\mu\nu} & \frac{\delta}{f_{\mu}^{a}(x)} \frac{\delta}{\delta f_{\nu}^{b}(x)} Z_{\lambda} [f] = \int d^{2}x\ \varepsilon^{ab} \varepsilon_{\mu\nu} \frac{\delta }{\delta f_{\mu}^{a}(x)} \int De\ \left( - \frac{1}{\lambda} \varepsilon_{bc} \varepsilon^{\nu\rho} (f-e)_{\rho}^{c}(x) \right) e^{- S_{K,\lambda}} Z_{0}[e] \nonumber\\
  &= - \frac{2}{\lambda} \delta(0) \left( \int d^{2}x \right) Z_{\lambda} + \int De\ \left( \frac{1}{2 \lambda^2} \int \varepsilon_{ab} (f-a)^{a} \wedge (f-e)^{b} \right) e^{-S_{K,\lambda}} Z_{0} [e].
  \label{eqn:flow-rhs}
\end{align}

We thus define the normal-ordering by subtracting out the piece proportional to $\delta(0)$. Note this term exists equally well in flat space.  With this prescription, the flow equation\eqref{eqn:flow-eqn} holds rather trivially:
\begin{align}
     \partial_{\lambda} \log Z_{\lambda}[f] & = \left\langle \frac{1}{ \lambda^2 }\int (f-e)^a \wedge (f-e)^b  \right\rangle \\
      & = \frac{1}{Z_{\lambda}[f]} \int d^2 x \varepsilon^{ab} \varepsilon_{\mu \nu } \frac{1}{2}  : \frac{\delta}{\delta f^{a}_{\mu}} \frac{\delta}{\delta f^{b}_{\nu}} : Z_{\lambda}[f] 
      \label{eqn:flow-equality}
\end{align}

Note, finally, that this proof did not care about the nature of the seed.
It can be a CFT, a QFT, or indeed a $T \bar{T}$-deformed theory itself.
It can also be a physically uninteresting function of the vielbein, but we will ignore this possibility.

\subsection{The ``Wheeler-de Witt'' Equation}
In this section, for completeness of presentation, we reproduce the main result of \cite{Freidel:2008sh}, namely that the kernel satisfies not just the flow equation but also a local equation, which in the 3d gravity interpretation is the Hamiltonian Wheeler-de Witt equation.

The equation is
\begin{align}
    \left\{ f_\mu^a (x) \delta_{f_\mu^a(x)} - \frac{c}{24\pi} (\det f) R[f](x) \right\} Z_0[f] &= 0 \nonumber\\
  \Rightarrow \left\{ f_{\mu}^{a}(x) \frac{\delta}{\delta f_{\mu}^{a}(x)} + \lambda \varepsilon^{ab} \varepsilon_{\mu\nu} :\frac{\delta}{\delta f_{\mu}^{a}(x)} \frac{\delta}{\delta f_{\nu}^{b}(x)}: - \frac{c-24}{24\pi} \det(f)R[f](x) \right\} Z_{\lambda} [f] &= 0.
  \label{eqn:wdw}
\end{align}
The first line is the requirement that the seed be a CFT.
For a non-CFT, there is a somewhat less tractable generalisation,
\begin{align}
    \left\{ f_\mu^a (x) \delta_{f_\mu^a(x)} - \frac{c}{24\pi} (\det f) R[f](x) \right\} Z^{QFT}_0[f] &= Z^{QFT}_0[f] \langle O(x) \rangle_{0,f} \nonumber\\
  \Rightarrow \left\{ f_{\mu}^{a}(x) \frac{\delta}{\delta f_{\mu}^{a}(x)} + \lambda \varepsilon^{ab} \varepsilon_{\mu\nu} :\frac{\delta}{\delta f_{\mu}^{a}(x)} \frac{\delta}{\delta f_{\nu}^{b}(x)}: - \frac{c-24}{24\pi} \det(f)R[f](x) \right\} Z^{QFT}_{\lambda} [f] &=  Z^{QFT}_{\lambda}[f] \left\langle \langle O(x) \rangle_{0,e} \right\rangle_{\lambda,f}.
  \label{eqn:wdw-qft}
\end{align}
On the right hand side of the first equality, $\langle O(x) \rangle_{0,f}$ stands for the undeformed field theory expectation value of the specific operators defining the QFT (away from a fixed point) multiplied by their respective beta functions. It is essentially a local version of the Callan-Symanzik equation (see \cite{Baume:2014rla} for a nice overview). The additional expectation value appearing on the RHS of the second line denotes the fact the field theory $\langle O(x)\rangle_{0,e}$ is then further averaged over geometries under the $e$-path integral. 
Crucially, the `normal ordering' in these two ``Wheeler-de Witt'' equations matches exactly \eqref{eqn:normal-ordering}, required for the flow equation to hold.

A further important point is that the flow \eqref{eqn:flow-eqn} and WdW equations \eqref{eqn:wdw} together give us a fully quantum version of the dimensional analysis equation \eqref{eqn:dim-anal-eqn}, with a shift in the effective central charge,
\begin{equation}
    \int \left\{ f_\mu^a \delta_{f_\mu^a} - \frac{c-24}{24\pi} (\det f) R[f] \right\} Z_\lambda [f] = - 2 \lambda \partial_\lambda Z_\lambda [f].
    \label{eqn:dim-anal-exact}
\end{equation}
There are again extra terms for a non-CFT seed. 
This is a non-trivial fact. It provides the underlying justification as to why the heuristic argument in the introduction makes any sense.
An equation of this form is usually expected to be true for relevant deformations. That it also holds for this irrelevant-seeming deformation is likely an important part of the solubility of $T \bar{T}$.



We will first derive \eqref{eqn:wdw} for a CFT, and then comment on the proof of \eqref{eqn:wdw-qft}.
First consider the action of the single derivative, scaling operator on the partition function:
\begin{align}
 f_{\mu}^{a}(x) \frac{\delta}{\delta f_{\mu}^{a} (x)} Z_{\lambda} [f] 
    &=  \int De\  \left( - \frac{1}{\lambda} f_{\mu}^{a} (x) \varepsilon^{\mu\nu} \varepsilon_{ab} (f-e)^{b}_{\nu} \right) e^{- S_{K}} Z_{0}[e] \nonumber\\
    &= \int De \left( - \frac{1}{\lambda} \varepsilon^{\mu\nu} \varepsilon_{ab} (f-e)_{\mu}^{a} (f-e)_{\nu}^{b} \right) e^{- S_{K}} Z_{0}[e] \nonumber\\
    &\qquad\qquad+ \int De \left( - \frac{1}{\lambda} \varepsilon^{\mu\nu} \varepsilon_{ab} e_{\mu}^{a} (f-e)_{\nu}^{b} \right)  e^{- S_{K}} Z_{0}[e] \nonumber\\
    &= - \lambda \varepsilon^{ab} \varepsilon_{\mu\nu} : \frac{\delta} {\delta f_{\mu}^{a}(x)} \frac{\delta}{ \delta f_{\nu}^{b} (x)} :  Z_{\lambda}[f] - \int De \left(  e_{\mu}^{a}(x) \frac{\delta}{\delta e_{\mu}^{a}(x)} e^{- S_{K}} \right) Z_{0}[e] \nonumber\\
  \label{eqn:wdw-check-poly-terms}
\end{align}
where in going to the final line, we used \eqref{eqn:flow-equality}. All that remains to be shown is 
\begin{equation}
    \int De \left( e_{\mu}^{a}(x) \frac{\delta}{\delta e_{\mu}^{a}(x)} e^{- S_{K}} \right) Z_{0}[e] = - \frac{c-24}{24 \pi} \det(f) R[f] Z_{\lambda}[f]
    \label{eqn:potential-piece}
\end{equation}

The proof of \cite{Freidel:2008sh} can be repackaged into two Schwinger-Dyson-like equations satisfied by the full path integral. 
One is for scalings of the base space vielbein $e$. The other concerns base space diffeomorphisms.

First, we notice that $e^{a}_{\mu} \delta_{e^{a}_{\mu }}$ generates local scale transformations. The basic idea is then to integrate by parts in \eqref{eqn:potential-piece}. Assuming no boundary contributions in field space, the scaling generator may then act on both the seed partition function, but also on the measure which transforms anomalously. More precisely, we can extract the dependence of the measure and seed partition function on the Weyl mode of the metric, $\Omega$, via a Liouville action 
\begin{equation}
    De Z_0[e]=  \hat{D}e \ e^{\frac{-24}{48 \pi } S_{L}} Z_{0}[\hat{e}]e^{+\frac{c}{48 \pi } S_{L}}
\end{equation}

In writing this equation, we have used the work of \cite{david1988conformal,distler1989conformal,d19902} which shows the exponentiated Liouville action provides the Jacobian between the Weyl anomalous measure $De$ and the scale invariant one $\hat{D}e$. We show in \ref{app:De} their results apply equally to well to translation-invariant measure we have chosen. We also elaborate on the origin of the coefficient -24 in the appendix. The scale dependence of the seed CFT partition function follows immediately upon integrating the conformal anomaly equation
\begin{equation}
e^{a}_{\mu} \frac{\delta}{\delta e^{a}_{\mu}} Z_{0}[e]= -\frac{c}{24 \pi } \det(e) R[e] Z_{0}[e].
\end{equation}

We can therefore write 
\begin{equation}
    \int De \left( e^{a}_{\mu}(x) \frac{\delta}{\delta e^{a}_{\mu}(x)} e^{-S_K} \right) Z_0[e] = -\int De \left(+ \frac{c-24}{24 \pi } \det(e) R[e] \right) e^{-S_K} Z_{0}[e].
       \label{eqn:WeylSD-CFT}
\end{equation}
This step, where we have used the conformal anomaly equation, is the only one that differs between CFTs and other QFTs.
For a QFT, this equation becomes instead
\begin{equation}
    \int De \left( e^{a}_{\mu}(x) \frac{\delta}{\delta e^{a}_{\mu}(x)} e^{-S_K} \right) Z_0[e] = -\int De \left(+ \frac{c-24}{24 \pi } \det(e) R[e] + \langle O \rangle_{0,e} \right) e^{-S_K} Z_{0}[e].
    \label{eqn:WeylSD-QFT}
\end{equation}
None of the following steps will differ between the two sides, and so finishing the proof of the CFT WdW will also furnish a proof of the more general one.

The curvature term appearing above is that relative to the dynamical vielbein $e$. The final step requires showing  $\det(f) R[f]=\langle \det(e) R[e] \rangle $. A Schwinger-Dyson type argument expressing diffeomorphism invariance provides the missing piece. Recall both the measure and seed partition  function are invariant under the transformation $e^a \rightarrow e^{(\xi) a}$ Infinitesimally, we may write $ e^{(\xi) a} = e^a + \grad Y^a + \mathcal{O}(Y^2 )$, where $\grad$ is the covariant derivative, $Y^a = i_{\xi}e^a$, and we have used $\grad e^a = 0$. 

Since $ De = D(e + \grad Y)$, the following path integral identity trivially holds
\begin{equation}
  \int De F[e + \grad Y] = \int D(e + \grad Y) F[e + \grad Y] = \int D \tilde{e} F[\tilde{e}].
  \label{eqn:int-invariance}
\end{equation}

Under such an infinitesimal transformation, 
\begin{equation}
  F[e + \grad Y ] = F[e] + \int d^{2}x \det(e) \grad_{\mu} Y^{a}(x) \frac{\delta}{\delta e_{\mu}^{a}(x)} F[e] 
\end{equation}
These two equations together imply\footnote{One further lesson here is that we can integrate $\grad_{\mu} \delta_{e_{\mu}^{a}}$ by parts in the $e$ integral, \emph{without any contact terms}.}
\begin{equation}
  \int De \int d^{2}x \grad_{\mu} Y^{a} \delta_{e_{\mu}^{a}} F[e] = - \int d^{2}x Y^{a} \int De \grad_{\mu} \delta_{e_{\mu}^{a}} F[e] = 0.
  \label{eqn:int-by-parts-eqn}
\end{equation}
In particular, using the diff-invariance of the seed $\grad_{\mu} \delta_{e_{\mu}^{a}} Z_{0}[e]=0$, we may write 
\begin{align}
  0 & = \int De \left( \grad_{\mu} \delta_{e_{\mu}^{a}} e^{- S_K} \right) Z_{0} [e] \label{eqn:kernel-diff-invariance} \nonumber \\
  & = \int De \left( \grad_{\mu} \varepsilon_{ac} \varepsilon^{\mu \rho } (e-f)^{c}_{\rho} \right) e^{- S_K} Z_{0}[e]\\
  \rightarrow 0 & =  df^a + \varepsilon^a_b \langle \omega[e] \rangle \wedge f^b
\end{align}

Since we can solve for the spin connection from the torsionlessness condition $df^a + \varepsilon^a_b \omega[f] \wedge f^b$, this implies that 
\begin{equation}
  \omega [f] = \langle \omega[e] \rangle + \text{no contact terms}.
  \label{eqn:omega-e-opf-new}
\end{equation}

Finally we can rewrite the Ricci scalar in terms of the spin connection as  $\varepsilon^{\mu \nu } \partial_{\mu } \omega[f]_{\nu}= \det(f) R[f]$. This concludes the proof that $Z_{\lambda}[f]$ solves the Wheeler-de-Witt equation, provided the $Z_{0}[f]$ solves the conformal anomaly equation.

\subsection{Intuitive Derivation of the Kernel}
\mn{V: So E and I have discussed this, but what I think we should use this section for is to say that one way to think of the kernel, and this is what Tolley did, is to say that it is the stochastic path integral that solves the diffusion like equation. But, what we are saying is that it is also related to the 3D gravity path integral which can be used to generate solutions to the WdW equation. }
\mn{R: copying Tolley seems dodgy. This is how we thought of it before Tolley; slightly less elegant but actually ours. That version is behind an iffalse if you guys feel strongly about it.}

This section attempts to provide an intuitive understanding for the form of the kernel. The recursive nature of the $T \bar{T}$ deformation begs the question as to simplicity of the kernel's $\lambda$-dependence. 

A fruitful perspective interprets the flow equation 
\begin{equation}
    \partial_{\lambda} Z_{\lambda}[f] = \frac{1}{2}\int_{\Sigma} d^2 x \varepsilon^{ab} \varepsilon_{\mu \nu }  :\frac{\delta} {\delta f_{\mu}^{a}(x)} \frac{\delta}{ \delta f_{\nu}^{b} (x)}:Z_{\lambda}[f]
\end{equation}
as a Schr\oe dinger-like equation for a wavefunction in the position basis $Z_{\lambda}[f]$, where the deformation parameter $\lambda$ plays the role of ``time'' and the Hamiltonian $\mathcal{H} \equiv \frac{1}{2} \int_{\Sigma} d^2 x \varepsilon^{ab} \varepsilon_{\mu \nu }  :\frac{\delta} {\delta f_{\mu}^{a}(x)} \frac{\delta}{ \delta f_{\nu}^{b} (x)}:$ is quadratic in the momentum operators. 

The main inspiration the argument below takes from the connection to 3D gravity is the identification of the deformed partition function with a radial wavefunction. Specifically, while recursive deformations like the $T \bar{T}$ deformation are somewhat novel \mn{is this too strong a statement? RG is recursive and pretty standard? Also, this is a totally different Hamiltonian... V: Indeed it's a different Hamiltonian but as far as the RG is concerned, what's special about this is not so much that it is recursive but that other irrelevant operators aren't triggered in subsequent iterations.}\mn{R: that's related to the recursive nature --- it's like picking the dirn of flow at every step.}, they are familiar as evolving wavefunctions.

The important observation to make is that above Hamiltonian $\mathcal{H}$ is itself independent of $\lambda$. We can therefore formally write down a solution to this flow equation as 

\begin{equation}
    Z_{\lambda}[f]= e^{-\frac{\lambda}{2}\int_{\Sigma} d^2 x \varepsilon^{ab} \varepsilon_{\mu \nu }  :\frac{\delta} {\delta f_{\mu}^{a}(x)} \frac{\delta}{ \delta f_{\nu}^{b} (x)}:} Z_{0}[f]
\end{equation}

As far as we know, this equation seems to have first appeared in \cite{cottrell2019comments}. To make this expression tractable, we should work in a basis which diagonalizes the Hamiltonian. 

As a warmup, consider the toy example from single-particle quantum mechanics.

\begin{equation}
  H = \hat{p}^{2} \quad \ket{\psi(t=0)}=\int dx \psi(x,0) \ket{x}
  \label{eqn:free-ham}
\end{equation}

we can straightforwardly write the time evolved state by working in the momentum basis: 

\begin{align}
    \ket{\psi(t)} & = e^{i t \hat{H} } \ket{\psi(t=0)} \nonumber \\
    & = e^{i t \hat{p}^2}\int \frac{dp}{2\pi} dx e^{- i p x } \psi(x,0) \ket{p} \\
    \rightarrow \psi(y,t) & = \int  \frac{dp}{2\pi} dx e^{- i p (x-y) + i p^2 t} \psi(x,0)  \propto \int dx e^{-\frac{i (x-y)^2}{4t}\psi(x,0) }
\end{align}

The representation of the deformed theory exactly analogous to this exponential form has been used successfully in \cite{Cardy:2018jho,Cardy:2018sdv} to understand many of its properties.
This parallel naturally leads on to ask what is the analog of momentum space for the $T \bar{T}$ deformation?

It turns out that this is a well-known object: the Legendre transformation that takes the partition function to exponential of the stress tensor effective action. This point of view was already advocated for in \cite{cottrell2019comments}. \mn{E: effective action woudl be $log Z[\pi] $  right? or is it defined as the Legrendre transform of $\log Z[e]$ in which case, what we're saying is wrong (log of fourier transform isn't same as fourier transform of log, we should just double check what the terminology is, i just don't know it}
\mn{V: Typically one does the Legendre transform of the generating functional for connected correlation functions and hence the log, whereas we are after the generating functional for all correlation functions, so what we are doing is the correct Legendre transform.}
This Legendre transformation may be written as 
\begin{equation}
   Z_{0}[\pi] = \int De \ e^{- \int \pi_{a} \wedge e^{a}} Z_{0} [e].
  \label{eqn:leg-trans-def}
\end{equation}
With this transform of the partition function, the $T \bar{T}$-deformed object is given merely by
\begin{equation}
  Z_{\lambda}[\pi] = e^{ \frac{\lambda}{2} \int \varepsilon_{ab} \pi^{a} \wedge \pi^{b}} Z_{0} [\pi].
  \label{eqn:deformed-eff-act}
\end{equation}
Performing the inverse Legendre transformation to obtain again a partition function, we find that \mn{E: there are also $\lambda$ dependent prefactors we have dropped in doing this integral. We shoudl say this as they could technically chancge flow equation. byt we can dop them via a renormalization of cosmological constant i think is the justification? You agree? V: yes, I think this is the how it should go.}
\begin{equation}
  Z_{\lambda} [f] = \int D\pi \ e^{\int \pi_{a} \wedge f^{a}} \Gamma_{\lambda} [\pi].
  \label{eqn:z-lamb-1}
\end{equation}
Finally, integrating out $\pi^{a}$ gives us
\begin{align}
  Z_{\lambda} [f] &= \int D\pi De \ e^{\int \pi_{a} \wedge (f^{a} - e^{a}) + \frac{\lambda}{2} \int \varepsilon^{ab} \pi_{a} \wedge \pi^{b}} Z_{0} [e] \nonumber\\
  &= \int De \ e^{- \frac{1}{2\lambda} \int \varepsilon_{ab} (f-e)^{a} \wedge (f-e)^{b}} Z_{0} [e].
  \label{eqn:z-lamb-derived}
\end{align}
This is exactly the Freidel kernel \eqref{eqn:kernel-intro}. \footnote{In going to the second line, we dropped a $\lambda$ dependent prefactor, which could naively affect the flow equation. This constant may be renormalized away via a cosmolomogical constant counter term \cite{polchinski1986evaluation}. }

In two dimensions, the above manipulations mainly serve to gain intuition about the deformation. In section \ref{3dgravity}, we show the classical equations of motion of the $\pi$ form of the Kernel give precisely the relation between Dirichlet boundary conditions at a cutoff surface and mixed boundary conditions at the asymptotic boundary of $AdS$ found in \cite{Guica:2019nzm}.

\subsection{Further Properties of the Kernel } \label{ssec:props}
In this section, we show some simple but important properties of the kernel. The first corresponds to the conservation and symmetry of the stress tensor, the second to the composition property of the integral kernel.

We first show the local stress tensor is conserved and symmetric by exhibiting the partition function is invariant under diffeomorphisms and local rotations,
\begin{align}
    \grad_\mu \langle T^\mu_a \rangle = 0 \quad &\Leftrightarrow \quad Z_\lambda [f] = Z_\lambda [f^\xi] \nonumber\\
    \varepsilon^a_{\ b} f_\mu^b \langle T^\mu_a \rangle = 0 \quad &\Leftrightarrow \quad Z_\lambda [f^a] = Z_\lambda [\left( e^{\phi \varepsilon} \right) ^{a}_{\ b} f^b].
    \label{eqn:t-props-as-symmetries}
\end{align}
It will be crucial for these proofs that the measure $De$ is invariant under both diffeomorphisms and local rotations.
Let us first show that the partition function is invariant under diffeomorphisms.
\begin{align}
  Z_\lambda[f^{(\xi)}] &=  \int De e^{- \frac{1}{\lambda} \int (f^{(\xi)}-e)^{1} \wedge (f^{(\xi)}-e)^{2}} Z_{0} [e] \nonumber\\
   &=  \int De^{(\xi)} e^{- \frac{1}{\lambda} \int (f^{(\xi)}-e^{(\xi)})^{1} \wedge (f^{(\xi)}-e^{(\xi)})^{2}} Z_{0} [e^{(\xi)}] \nonumber\\
   &=  \int De e^{- \frac{1}{\lambda} \int (f-e)^{1} \wedge (f-e)^{2}} Z_{0} [e] \nonumber \\
\rightarrow Z_\lambda[f^{(\xi)}] &= Z_\lambda[f]
  \label{eqn:diffinvariance}
\end{align}
where in going to the second line, we simply \emph{relabeled} the integration variable $e \rightarrow e^{\xi}$. The third equality, however, requires diffeomorphism invariance of the measure $De=De^{(\xi)}$ and the action (what appears in the exponents), along with the fact that $Z_{0}[e]=Z_{0}[e^{(\xi)}]$.\footnote{ This "trick" might be most familiar from the Fadeev-Poppov procedure used for gauge theory path integrals. } The second line in \eqref{eqn:t-props-as-symmetries} can be shown similarly, using rotational invariance of the measure.

Next, we show that the kernel composes, that is
\begin{equation}
    \int De \ e^{- \frac{1}{2\lambda_2} \int \varepsilon_{ab} (f-e)^a \wedge (f-e)^b} Z_{\lambda_1} [e] = Z_{\lambda_1 + \lambda_2} [f].
    \label{eqn:composition}
\end{equation}
The $e$-dependent part of the action in \eqref{eqn:composition} is
\begin{equation}
    S_{K,\lambda_2} [f,e] + S_{K,\lambda_1} [e,\tilde{e}] = \frac{1}{2 \lambda_2} \int \varepsilon_{ab} (f-e)^a \wedge (f-e)^b + \frac{1}{2 \lambda_1} \int \varepsilon_{ab} (e - \tilde{e})^a \wedge (e - \tilde{e})^b.
    \label{eqn:e-part}
\end{equation}
We see that the dependence on the intermediate vielbein $e$ is quadratic; in gravitational path integrals, this does not always mean that one can integrate it out in the usual way.
However, because of the existence of a translation-invariant measure, in this case one can.\footnote{Yet again, we thank A. Tolley for pointing out the existence of this measure to us; this was a bit of a puzzle.}
One then finds, as expected,
\begin{equation}
    S_{K,\lambda_1 + \lambda_2} [f,\tilde{e}] = \frac{1}{2(\lambda_1 + \lambda_2)} \int \varepsilon_{ab} (f-e)^a \wedge (f-e)^b.
    \label{eqn:composed-kernel}
\end{equation}
This also explains why this form of the deformation looks like the simplest way to ``exponentiate'' the Cardy argument, which gives for an infinitesimal deformation
\begin{equation}
    Z_{\lambda + \delta \lambda} [f] = \int D \delta e\ e^{- \frac{1}{2 \delta \lambda} \int \varepsilon_{ab} \delta e^1 \wedge \delta e^2} Z_\lambda [f-\delta e].
    \label{eqn:cardy}
\end{equation}
Since this form composes, it naturally retains its form at finite deformation as well.

\subsection{Spatially Varying Deformations}
One generalisation comes from merely increasing the number of flow equations to match the number of WdW equations.
Allowing $\lambda$ to vary spatially, $\lambda \rightarrow \lambda(x)$. the kernel becomes
\begin{equation}
    Z_{\lambda,x}[f]= \int De \ e^{-\int \frac{1}{2\lambda(x)} \varepsilon_{ab}(f-e)^{a}  \wedge (f-e)^b } Z_{0}[e].
\end{equation}
By the same logic as above, it solves the local flow equation
\begin{align}
     \frac{\delta}{\delta \lambda(x)} Z_{\lambda,x} [f] &=  \left( \frac{1}{2} \varepsilon^{ab} \varepsilon_{\mu\nu} :\frac{\delta}{\delta f_{\mu}^{a}(x)} \frac{\delta}{\delta f_{\nu}^{b}(x)}: \right) Z_{\lambda,x} [f].
     \label{eqn:local-flow-eqn}
\end{align}
The WdW equation is no longer as simple, because of the spatial dependence of $\lambda$.


Indeed, the proof of the WdW equation proceeds similarly, except that the diffeomorphism Schwinger-Dyson equation now gives
\begin{align}
    0 &= Z_\lambda^{-1} \int De \grad_\mu \delta_{e_\mu^a} e^{-S_K} Z_0 [e] \nonumber\\
    &= \left\langle \varepsilon_{ab} \varepsilon^{\mu\nu} \grad_\mu \frac{(e-f)^b_{\nu}}{\lambda(x)} \right\rangle \nonumber\\
    &= \varepsilon_{ab} \varepsilon^{\mu\nu} \left\langle \grad_\mu f_\nu^b +\frac{(e-f)_\nu^b}{\lambda(x)} \partial_{\mu}  \lambda(x) \right\rangle
    \label{eqn:mod-SD-diffs}
\end{align}

An interesting point about this deformation is that the space-time dependence of $\lambda$ causes the deformed stress tensor to not be conserved.

Straightforward but tedious algebra - which we include in \ref{app:modWDW} - shows that \eqref{eqn:mod-SD-diffs} implies

 \begin{equation}
     \langle \det(e) R[e] \rangle Z_{\lambda }[f]= \det(f) R[f] Z_{\lambda}[f] + \varepsilon^{\alpha \sigma} \partial_{\alpha} \left( \frac{1}{\det(f)}f_{a,\sigma} \varepsilon^{ac} (\partial_{\nu}\lambda) \frac{\delta}{\delta f^{c}_{\nu}(x)} \right) Z_{\lambda}[f]
 \end{equation}
 
All the manipulations in \eqref{eqn:wdw-check-poly-terms} still hold, with $\lambda \rightarrow \lambda(x)$ as do \eqref{eqn:WeylSD-CFT} and \eqref{eqn:WeylSD-QFT}.  Therefore, the only change to \eqref{eqn:wdw} and \eqref{eqn:wdw-qft} is to shift the curvature piece as 

\begin{equation}
    \det(f) R[f] Z_{\lambda}[f] \rightarrow \det(f) R[f] Z_{\lambda}[f] + \varepsilon^{\alpha \sigma} \partial_{\alpha} \left( \frac{1}{\det(f)}f_{a,\sigma} \varepsilon^{ac} (\partial_{\nu}\lambda) \frac{\delta}{\delta f^{c}_{\nu}(x)} \right) Z_{\lambda}[f]
\end{equation}

\mn{E to E : add Wdw modification here ASAP}

\subsection{Relation to Flat Space ``JT'-Gravity'' Proposal } \label{ssec:dghc}
The beautiful work of Dubovsky et al. \cite{Dubovsky:2012wk,Dubovsky:2017cnj,Dubovsky:2018bmo} provided the impetus for our own. The authors of \cite{Dubovsky:2018bmo} in particular, whose initials abbreviate to DGHC, precisely recast the $T\bar{T}$ deformation in flat space as the coupling of the undeformed theory to 2d topological gravity. They explicitly computed the torus path integral. It reproduced the dressing of the energy levels known from solving the relevant inviscid Burger's equation. 
The narrative there focused around a variant of Jackiw-Teitelboim gravity, dubbed JT', with an action of the form 
\begin{equation}
    S_{JT'}[g,\phi]]=\int d^2 x \sqrt{g} \left( \phi R - \Lambda \right)
\end{equation}

The dilaton $\phi$ serves as Lagrange multiplier to ensure $R=0$. The cosmological constant term is taken to be inversely proportional to the $T \bar{T}$ deformation parameter. Switching to a first order formalism requires Lagrange multipliers enforcing the torsionlessness constraints, leading to the action 
\begin{equation}
    S_{JT'}[e,\omega,\mu,\phi]= \int \phi d\omega -  \frac{\Lambda}{2}\varepsilon_{ab}e^a \wedge e^b +  \mu_{a} \left( de^a + \omega \epsilon^{a}_{b} \wedge e^b \right) 
\end{equation}

The important insight was the Lagrange multipliers could be viewed as providing a map from the base space to the target space on which the deformed theory lived. However, to arrive at the final form of the path integral they computed, they needed i) to neglect any possible holonomies of the spin connection on the torus (the dilaton sets $d\omega =0 $, but not $\omega=0$), ii) supplement the action by an additional term to match the deformed energies,\footnote{In their notation, a term proportional to $\int \varepsilon_{ab} dX^a \wedge dX^b$, which does vanish as total derivative since $X$ winds.} and iii) limit the field range of the Lagrange multiplier zero modes giving rise to an important factor of the target space torus' area. This suggests the coupling to  Jackiw-Teitelboim gravity is not the actual root of the DGHC integral kernel.\footnote{We should point out that we are by no means the first people to notice this fact, see for example \cite{VictorsTalk}} 

The better way to think of it is as a particular form of the kernel.
We showed in section \ref{ssec:props} that $Z_{\lambda}[f]=Z_{\lambda}[f^{(\xi)}]$; in other words, that the deformed theory's partition function is invariant under target space diffeomorphisms (generated by the vector field $\xi$). We can therefore average $Z_{\lambda}[f]$ over all target space diffeomorphism as long as we also divide out by the volume of diffeomorphism group 
\begin{align}
     Z_{\lambda}[f] & =\left(1=\int \frac{D\xi }{\text{Vol(Diff)}} \right) Z_{\lambda}[f] = \int \frac{D\xi }{\text{Vol(Diff)}} Z_{\lambda}[f^\xi].
     \label{eqn:ts-int-kernel}
\end{align}


When $df^a=0$, or alternatively $\omega_{(f)}=0$, as in DGHC, diffeormorphisms are rather simple. Under a general coordinate transformation $x^{\mu} \rightarrow \tilde{x}^{\mu}(x) $, the vielbein transforms as 

\begin{equation}
    f^a_{\mu}(x) \rightarrow \frac{\partial x^{\nu}(y)}{\partial \tilde{x}^{\mu} }f_{\nu }^a(x(\tilde{x})) \overset{f \text{constant}}{=}  \frac{\partial x^{\nu}(y)}{\partial \tilde{x}^{\mu} }  f_{\nu }^a.
\end{equation}


Hence, while this may at first look like an infinitesimal version of a diffeomorphism, we maybe in fact alternatively parameterize the full $f^{a (\xi)}$ as $f^a+dY^a$. This $Y^a$ are related to the full diffeomorphism via  $\tilde{x}^{\mu}(x)=x^{\mu}+f^{\mu}_{a}Y^{a}(x)$. We may then rewrite \eqref{eqn:ts-int-kernel} as 
\begin{equation}
    Z_{\lambda}[f]= \int \frac{DY De}{\text{Vol(Diff)}} e^{-\frac{1}{\lambda}\int \varepsilon_{ab} (f+dY-e)^{a} \wedge (f+dY-e)^{b}} Z_{0}[e]
    \label{eqn:DGHCresult}
\end{equation}
which is none other than the DGHC kernel (cf. their Eqn. 28, with $dX^a=f^a+dY^a$ ).\footnote{Technically, their measure $De_{\tiny{DGHC}}$ differs from ours. In appendix \ref{app:e}, we show how the two ultimately give rise to the same path integral.}

One might worry that what we have introduced in the denominator is not the volume of the base space diffeomorphisms but those of the target space; further, the measure for the integral over target space diffeomorphisms in the numerator is defined with respect to the target space metric. 
It turns out that these two subtleties cancel out.
There are two ways the measures can have metric dependence: the range, and the inner product the measure is defined with respect to.
Since the $\xi$s are coordinate redefinitions, their range is defined by the coordinate range and so doesn't depend on the metric.
The measures in \eqref{eqn:ts-int-kernel}, however, are defined with respect to the inner product,
\begin{equation}
    (\delta \xi, \delta \xi) = \int (\det f) g_{(f)\mu\nu} \delta \xi^\mu \delta \xi^\nu.
    \label{eqn:ts-diff-ip}
\end{equation}
However, this dependence cancels between the numerator and the denominator.
Any one form in two dimensions can be written in terms of two scalars and zero-modes valued in the first homology group of the manifold,
\begin{equation}
    \xi = d\alpha + *d\beta + \bar{\xi}, \quad \bar{\xi} \in H^1.
    \label{eqn:one-form-decomp}
\end{equation}
For the non-zero-modes that can be parametrised by the scalars, it is shown in appendix \ref{app:De} that there is no metric dependence leftover.
For the zero-modes, the metric dependence cancels between the numerator and denominator.
Thus, we may as well write both integrals with respect to the base space and recover the flat space kernel of \cite{Dubovsky:2017cnj,Dubovsky:2018bmo}. This provides an important check on our proposal. It shows that $Z_{\lambda}[f]$ gives the right answer for the case of the torus. In particular, our kernel therefore reproduces the correct dressing of the energy levels known otherwise from solving the inviscid Burger's equation. 

\mn{Shall we write " Finally, to compare with \eqref{eqn:DGHCresult}, note that the range of the $Y^a$ integral giving an important area of the target space torus follows from from the relation $Y^a=f^{a}_{\mu}\xi^{\mu}$". }

\section{The 3D Story: Radial Wheeler-de Witt Wavefunctions} \label{3dgravity}
In this section, we explore the deep connection between $T\bar{T}$ and 3d gravity. The crux of the story is that solutions of a modified Wheeler-de Witt equation also furnish solutions to the $T \bar{T}$ flow equation.\footnote{The other way is not obvious at all.} This equation is best viewed as the Schr\oe dinger equation in a 2+1 decomposition of the gravity theory. The wavefunction depends on the metric on each two-dimensional slice in the same way the $T\bar{T}$ partition function depends on the geometry on which it is defined. 

We give a high-level overview of the difference in interpretation between the 3d and 2d understandings of the same objects in this table.

\vspace{0.5 cm}
\begin{tabular}{|c|c|}
\hline 
2d Object & 3d Gravity Interpretation\tabularnewline
\hline 
\hline 
Partition function $Z_{\lambda}[f]$ & Radial Wavefunction$\Psi_{WdW}[f]$\tabularnewline
Trace Flow equation & Zero-mode of Hamiltonian WdW equation \tabularnewline
Deformation kernel & Annular path integral \tabularnewline
Expectation value of stress tensor & Action of operator conjugate to metric \tabularnewline
- & Expectation value of operators \tabularnewline
Global diff and Lorentz symmetries & Gauge constraints \tabularnewline
Legendre Transform & Change of basis \tabularnewline

\hline

\end{tabular}

\vspace{0.5 cm}

\subsection{2+1 Decomposition in First-Order Formalism}
To keep this paper rather self-contained, we briefly review the 2+1 decomposition of 3d gravity in first-order formalism. The brief section V of \cite{Freidel:2008sh} informed much of the discussion below. We also found \cite{Carlip:2004ba} helpful for a more detailed treatment.

Instead of the second-order metric variables, the local frame fields or vielbeins form the fundamental degrees of freedom. We explain how this quantum mechanical problem reduces to a system of constraints which the wavefunction must satisfy. Each constraint has a clear geometrical interpretation. The Hamiltonian Wheeler-de-Witt equation is often also called the refoliation constraint. It points out the arbitrariness in our 2+1 split of the 3d geometry and dictates how the wavefunction must transform under different slicings of the full geometry. The two other relevant constraints correspond to diffeomorphism invariance on the 2d slice and, because we are using first-order variables, the freedom to make local 2d Lorentz transformations (rotations in our Euclidean setup).  

We can write the 3d metric in terms of the vielbeins via the standard relation
\begin{equation}
    g^{(3)}_{\alpha \beta }(x)=\delta_{ij}E_{\alpha}^{i}(x) E_{\beta}^{j}(x)
\end{equation}
The spin connection is defined as  
\begin{equation}
    dE^i+\omega^{i}_{\ j} \wedge E^{j}=0
\end{equation}
A peculiarity of 3d is that we may define a one indexed spin connection using the Levi-Civita symbol $\omega^{i}=\varepsilon^{ijk} \omega_{jk}$. In terms of these variables, the Einstein Hilbert action for a 3d spacetime with negative cosmological constant reads (after a rescaling of the vielbeins): 
\begin{equation}
  S_{3d} = \frac{l}{16\pi G_{N}} \int_{\mathcal{M}_3}  E_{i} \wedge R(\omega)^{i} - \frac{1}{6} \varepsilon_{ijk} E^{i} \wedge E^{j} \wedge E^{k}, \quad R^{i} \equiv d\omega^{i} + \varepsilon^{ijk} \omega_{j} \wedge \omega_{k},
  \label{eqn:3d-S}
\end{equation}
where $G_{N}$ is Newton's constant and $l$ is the $AdS_3$ radius of curvature. \mn{laurent has 1/8pi G , we OK with this normalization here?}
 
Looking towards a Hamiltonian analysis, consider now a foliation of the 3d geometry by 2d submanifolds (the equal ``time'' slices of canonical quantisation), $\mathcal{M}_3= \Sigma \times \mathbb{R}$. Using a locally adapted coordinate system with normal direction labeled by a coordinate $r$ and coordinates $x^{\mu}$ on the 2d slice, we decompose the vielbeins and spin connections as:
\begin{align}
  E^{0} &= E^{0}_{r}dr + n_{\mu} dx^{\mu}  \nonumber\\
  \omega^{0} &= \omega^{0}_{r}dr + \omega_{\mu} dx^{\mu} \nonumber\\
   E^{a} &= E^{a}_{r}dr + f^{a}_{\mu} dx^{\mu}  \nonumber\\
   \omega^{a} &= \omega^{a}_{r} dr + \pi^{a}_{\mu} dx^{\mu}  .
  \label{eqn:2d-names}
\end{align}
In terms of these, the action becomes
\begin{align}
  S_{3d} = \frac{l}{16\pi G_{N}} \int dr \int_{\Sigma} \ n \wedge \dot{\omega} + f^{a} \wedge \dot{\pi}_{a} &+ E_{r,0} \left\{ d \omega + \frac{1}{2} \varepsilon_{ab} \left( \pi^{a} \wedge \pi^{b} -  f^{a} \wedge f^{b} \right) \right\} \nonumber\\
  & +  E_{r}^{a} \left\{ d \pi_{a} - \varepsilon_{ab} \left( \omega \wedge \pi^{b} - n \wedge f^{b} \right) \right\} \nonumber\\
  & + \omega_{r}^{0} \left\{ dn - \varepsilon_{ab} \pi^{a} \wedge f^{b} \right\} \nonumber\\
  & + \omega_{r}^{a} \left\{ df^{a} + \varepsilon_{ab} \left( \omega \wedge f^{b} - \pi^{b} \wedge n \right) \right\}.
  \label{eqn:3d-s-decomposed}
\end{align}
where the dot denotes the partial derivative with respect to the radial coordinate. We may view this as a Hamiltonian system with canonically conjugate variables $\{n_{\mu},\varepsilon^{\mu \nu}\omega_{\nu} \}$ and $\{f^{a}_{\mu},\varepsilon^{\mu \nu }\pi_{a,\nu} \}$. \mn{agree with this def of momentum here? because of the wedge} The radial components of the vielbeins and of 3d spin connection serve as Lagrange multipliers enforcing constraints. From this form of the action, we see the Hamiltonian consists solely of these constraints $C_{\alpha}$, which we label as 
\begin{align}
  H &= d\omega + \frac{1}{2} \varepsilon_{ab} \left(\pi^{a} \wedge \pi^{b} - f^{a} \wedge f^{b}\right) \nonumber\\
  P_{a} &= d\pi_{a} - \varepsilon_{ab} (\omega \wedge \pi^{b} - n \wedge f^{b} ) \nonumber\\
  G &= dn - \varepsilon_{ab} \pi^{a} \wedge f^{b} \nonumber\\
  G_{a} &= df_{a} + \varepsilon_{ab} (\omega \wedge f^{b} + n \wedge \pi^{b}).
  \label{eqn:constraints}
\end{align}

Note these constraints  $C_{\alpha}$ are local, and hold pointwise on $\Sigma$. The $H$ constraint encodes invariance under refoliations. Its quantisation leads to the Wheeler-de-Witt equation.  The two $P_a$ constraints correspond to diffeomorphisms tangential to the 2d surface while $G$ generate local Lorentz rotations.

As can be seen from \eqref{eqn:2d-names}, the induced metric on $\Sigma$ is 

\begin{equation}
    ds^2_{\Sigma}=\left( \delta_{ab}f^{a}_{\mu}f^{b}_{\nu}+n_{\mu}n_{\nu} \right) dx^\mu dx^{\nu}
\end{equation}.
To make contact with the second-order formalism, we can use the $G_a$ constraints to set the redundant variable $n_\mu$ to zero; this amounts to orienting the local tangent spaces to agree with the foliation. Following \cite{Freidel:2008sh}, we call this `radial' gauge.
In this gauge the $G_a$ constraints just become the torsionlessness constraint setting $\omega$ to be a function of the vielbeins.

To quantise, we promote the momenta $\pi_{a,\mu}$ to the differential operators 
\begin{equation}
    \hat{\pi}(x)_{a,\mu}=\frac{16 \pi G_{N}}{l} \varepsilon_{\mu \nu} \frac{\delta}{\delta f(x)^{a}_{\nu}}
\end{equation}
At this point, one should be worried about ordering ambiguities in \eqref{eqn:constraints}, as well the coincident limit of the double derivative appearing in $\varepsilon_{ab} \pi^{a} \wedge \pi^{b}$. We will see the form of the kernel deals with these issues in rather subtle but remarkable ways, and hence temporarily postpone a more detailed discussion. 

In radial gauge, the wave-functional $\Psi_{WdW}[f]$ depends solely on the 2d vielbeins $f^{a}_{\mu}$ on each slice $\Sigma$ and must satisfy the local constraints.

\begin{equation}
    \hat{H}(x)\Psi_{WdW}=\hat{P}_{a}(x) \Psi_{WdW}=\hat{G}(x) \Psi_{WdW}=0
\end{equation}

\subsection{Wheeler-de-Witt Wavefunctional and the Kernel}

The remarkable result of \cite{Freidel:2008sh}, motivated by early work of Verlinde \cite{verlinde1990conformal}, maps the partition function of \emph{any} CFT onto a Wheeler-de-Witt wavefunctional in the radial gauge via the following integral transform: 
\begin{equation}
  \Psi_{WdW} [f] = e^{\frac{l}{16\pi G_{N}} \int f^1 \wedge f^2} \int De e^{- \frac{l}{8\pi G_{N}} \int (f-e)^{1} \wedge (f-e)^{2}} Z_{0} [e].
  \label{eqn:laurent-answer}
\end{equation}
where $Z_{0}[e]$ denotes the partition function of a CFT living on a background with metric $ds^2=\delta_{ab}e_{\mu}^{a}e_{\mu}^{b}dx^{\mu} dx^{\nu}$.

$\Psi_{WdW}[f]$ satisfies all the constraints discussed above. First, since the $P_a$ constraints are generators of 2d tangential diffeomorphisms, $\hat{P}_{a} \Psi_{WdW}=0$ is simply the infinitesimal statement that $\Psi_{WdW}[f]$ depends only on the diffeomorphism invariant data of $f$, i.e.
\begin{equation}
    \Psi_{WdW}[f]=\Psi_{WdW}[f^{(\xi)}]
\end{equation}
where $f^{(\xi)}$ denotes a diffeomorphism of $f$ (infinitesimally, $f^{(\xi)}=f+\mathcal{L}^{(D)}_{\xi}f = f+ D(i_{\xi}f)$.\footnote{The use of a ``gauge covariant'' Lie derivative here, which we denote by $\mathcal{L}^{(D)}_{\xi}$, rests upon the fact that we may also perform a compensatory lotal rotation generated by the spin connection. Using the spin connection compatible with $f$, i.e. satisfying $Df=0$, we find $\mathcal{L}^{(D)}_{\xi}f= D (i_{\xi}f) + i_{\xi}Df = D (i_{\xi}f) $ }
This was shown in section \ref{2d} in \eqref{eqn:diffinvariance}.
The $H$ constraint takes center stage in the relation to $T\bar{T}$. Commuting it past the first exponential in \eqref{eqn:laurent-answer} shows how the trace flow equations appears. Indeed, using 
\begin{equation}
    \hat{\pi}_{a \mu } e^{\frac{l}{16\pi G_{N}} \int f^{1} \wedge f^{2}} = e^{\frac{l}{16\pi G_{N}} \int f^{1} \wedge f^{2}} \left(  \hat{\pi}_{a \mu } - \varepsilon_{ab}f^{b}_{\mu} \right)
\end{equation}
we find that
\begin{equation}
    \hat{H} \Psi_{WdW}[f]= e^{\frac{l}{16\pi G_{N}} \int f^{1} \wedge f^{2}} \left( d\omega_{f} + \frac{1}{2}\varepsilon^{ab} \hat{\pi}_{a} \wedge \hat{\pi}_b + \frac{1}{2} \hat{\pi}_{a} \wedge f^a -  \frac{1}{2}   f^a \wedge \hat{\pi}_{a} \right) \int De \ e^{- \frac{l}{8\pi G_{N}} \int (f-e)^{1} \wedge (f-e)^{2}} Z_{0} [e]
\end{equation}

\mn{shall we rescale stuff here already to have $G_N$ and $l$ in the equation?}
This equation merits two observations. First off, we see the trace flow equation appearing, with $ \frac{1}{2}\epsilon_{ab} \hat{\pi}^{a} \wedge \hat{\pi}^b$ playing the role of $T\bar{T}$, $d\omega_f$ is simply the (hodge dual of) 2d Ricci scalar, and finally $\frac{1}{2} \hat{\pi}_{a} \wedge f^a -  \frac{1}{2}   f^a \wedge \hat{\pi}_{a}$ the dilation operator generating $tr(T)$, the trace of the deformed theory's stress tensor, up to a contact term.
In fact, this form of the refoliation constraint appeared long ago, in early work on holographic RG. It can be understood as a renormalization of the WdW equation using the Balasubramanian-Kraus (BK) counterterm and identifying the stress tensor with the Brown-York stress tensor on the slice. \cite{McGough:2016lol} already highlighted its relevance for $T \bar{T}$. \mn{(Vasu, I am sure your papers did too, so add them here! :) OTHER PAPERS, PEOPLE WILL GET ANGRY HERE, LETS CITE PROFUSELY }

Secondly, we see the $WdW$ equation even prescribes a particular ordering of the dilatation operator. The divergence arising from $\pi_a$ acting on $f_a$ in that expression is exactly canceled by the coincident divergence in the double functional derivative of $\varepsilon_{ab} \hat{\pi}^{a} \wedge \hat{\pi}^b$.
In other words, the reason we had to `normal order' the double derivative in the WdW equation in section \ref{2d} is that we wrote the trace term simply as $\pi_a \wedge f^a$.

To match this with our previous expressions for the the Wheeler-de Witt, we need the gravitational constant, AdS radius, and central charge to be related by
\begin{equation}
    c = \frac{3l}{2 G_N} - 12.
    \label{eqn:c-l-G-reln}
\end{equation}
This is the usual holographic relation, but with a shift because of the conformal anomaly of the measure. This shift is not worrisome, since our point of view is that the connection to holography is only an approximate large $c$ feature of this 3d gravity theory.

\mn{isn't there a renormalization of Brown-Henneaux formula even for usual holography? I'm actually not 100 per cent sure of this (the thing i was thinking about was this schwarzian for Ads3 but that's not holography, but just the edge theory of 3d CS with gravity boundary conditions }

Further, we can go to the ``$T \bar{T}$ version'' of the kernel \eqref{eqn:laurent-kernel} by the identification
\begin{equation}
    \frac{\lambda}{r^2} = \frac{8\pi G_N}{l},
    \label{eqn:lambda-1}
\end{equation}
where $r$ is some characteristic scale of the Dirichlet wall.
Rescaling $f$ and concomitantly $e$ by $r$, we find that the wavefunction becomes
\begin{equation}
    \Psi_{WdW} [r f] = e^{\frac{1}{2\lambda} \int f^1 \wedge f^2} r^{\frac{c-24}{3}} \int De e^{- \frac{1}{\lambda} \int (f-e)^1 \wedge (f-e)^2} Z_0 [e].
    \label{eqn:psi-lambda-1}
\end{equation}
We can get rid of the prefactors outside the integral by cosmological constant and curvature counterterms respectively to obtain the deformed partition function.

Finally, this derivation also makes sense of the existence of the `dimensional analysis' equation \eqref{eqn:dim-anal-exact}.
Since neither $\lambda$ not $r$ have an independent definition, one can rescale both by a concomitant amount, and only pay a cost in a new curvature counterterm of the form in \eqref{eqn:psi-lambda-1}.
Further, this dimensional analysis equation is the basis for converting the WdW equation into the flow equation.
This also explains why the WdW equation for the spatially varying $\lambda$ case is less nice --- in that case, one has to also let $r$ vary and obtain a full Liouville outside that one forgets about; passing the WdW through this creates extra terms.
The main physical output of this discussion is that the flow equation can be thought of as an epiphenomenon of looking at the WdW equation in `bad' variables.

\subsection{Interpretation as 3d gravity annular path integral}

We pursue our geometrization of the results of Section \ref{2d}. So far, we discussed how the deformed theory's partition function $Z_{\lambda}[f]$ solves all the constraint equations arising from a canonical treatment of 3d quantum gravity. The wavefunction $Z_{\lambda}[f]$, being a functional of the vielbein $f$ on the 2d slice, is expressed in the analog of the position basis \footnote{Dirichlet boundary conditions in quantum gravity are notoriously tricky. We will mostly gloss over those subtleties in this discussion. }

\begin{equation}
    Z_{\lambda}[f]=\bra{f}\ket{\Psi_{\lambda}}
\end{equation}

Satisfying $\hat{C}_{\alpha} \ket{\Psi_{\lambda}}=0$, the state $\ket{\Psi_{\lambda}}$ resides in the physical Hilbert space. It is also convenient to write it in the analog of the momentum basis

\begin{equation}
    \ket{\Psi_{\lambda}}= \int D\pi \ \Psi_{\lambda}[\pi] \ket{\pi}= \int D\pi \left( \int De e^{\int_{\Sigma }\frac{\lambda}{2}\varepsilon^{ab} \pi_{a}\wedge \pi_{b}-\pi_a \wedge e^a} Z_{0}[e] \right) \ket{\pi}
\end{equation}

where we used the overlap $\bra{\pi}\ket{e}=e^{-\int_{\Sigma} \pi_{a} \wedge e^a}$.

We stress the state $\ket{\beta}=\int De Z_{0}[e] \ket{e}$, built purely from the CFT partition function, does not solve the second order Wheeeler-de Witt equation (but rather "only" the first order conformal anomaly equation). This makes clear that even an infinitesimal deformation radically alters the nature of the state. 

An annular path integral corresponds to a transition amplitude. The two sets of boundary conditions on either side encode the initial and final state data. In 3d gravity, transition amplitudes involving at least one physical state reduce to an overlap. Indeed, since the  total Hamiltonian is simply the sum of constraints $C_{\alpha}$, which annihilate any physical state, we have 

\begin{equation}
    \bra{\phi} e^{-s \hat{H}_{tot}} \ket{\Psi_{\text{phys}}}= \bra{\phi} e^{-s \sum_{\alpha} \hat{C_{\alpha}}} \ket{\Psi_{\text{phys}}} = \bra{\phi}\ket{\Psi_{\text{phys}}}
\end{equation}

\begin{figure}[h] 
\label{pathintfig}
\centering
\includegraphics[width=8cm]{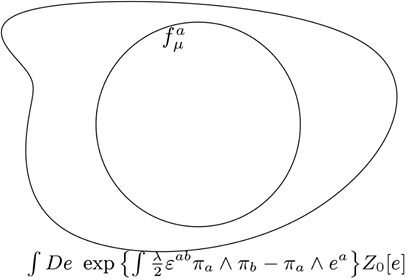}
\caption{$Z_{\lambda}[f]$ may be viewed as a transition amplitude between the states $\ket{\Psi_{\lambda}}=\int D\pi \left( \int De e^{\int_{\Sigma }\frac{\lambda}{2}\varepsilon^{ab} \pi_{a}\wedge \pi_{b}-\pi_a \wedge e^a} Z_{0}[e] \right) \ket{\pi}$ and $\ket{f}$. Geometrically, this corresponds to a 3d gravity path integral on an annulus, with particular choice of mixed boundary conditions on the outer edge and Dirichlet ones for the inner edge.}
\end{figure}

We may thus equally view the partition function $Z_{\lambda}[f]$ as a transition amplitude between the state $\ket{f}$, corresponding to fixing the vielbein on one of the 2d boundaries, and the state $\ket{\Psi_{\lambda}}$ on the other. See figure \ref{pathintfig}.

Let us spell out the connection to the 3d gravity path integral and the choice of boundary conditions a bit further. We will show how a semi-classical treatment of our asymptotic boundary state connects to the boundary conditions discussed in \cite{Guica:2019nzm}. Inserting a resolution of the identity, we can write 

\begin{align}
    Z_{\lambda}[f] & = \int De \bra{f}e^{-s\hat{H}_{tot}} \ket{e}\Psi_{\lambda}[e] \nonumber \\ 
    & = \int De \bra{f}e^{-s\hat{H}_{tot}} \ket{e} \left( \int D\pi e^{-\int \pi_a \wedge e^a + \frac{\lambda}{2} \varepsilon^{ab} \pi_a \wedge \pi_b -\log Z_{0}[e]} \right)
\end{align}

The 3d gravity path integral of interest is schematically then:


\begin{equation}
Z_{\lambda}[f]=\int De \Psi_{\lambda}[e] \int^{E|_{\infty}=e}_{E|_{r_{c}}=f} D E\, e^{-S_{GR}(E)}
\label{eqn:3dint}
\end{equation}

\mn{This equation can't even make sense as is. $E^i$ has three tangent space indices, $e^a$ only two. Need to include and $E^0$ being set to zero here too i think. }
\mn{R: It's okay they're Lagrange multipliers --- physical states are effectively at $\pi_{Lag mult} = 0$. But you can also set the Lag mults to 0 and it won't make a difference to the evolution of the physical states. Don't particularly need to go into it unless you feel like it.}
Here $S_{GR}(E)$ is the action for three dimensional gravity in first order variables with a negative cosmological constant along with the appropriate boundary terms needed for finiteness and for a well posed variational principle. The integration variable $E$ is the bulk vielbein and we gauge fix to the $n=0$ gauge. The boundary values are related to the vielbeins $e$ and $f$ that feature in the kernel. 

In the classical limit, we may evaluate $Z_{\lambda}[f]$ via a steepest descent approximation: 
\begin{equation}
Z_{\lambda}[f] \rightarrow e^{-\int \bar{\pi}_a \wedge \bar{e}^a - \frac{\lambda}{2} \varepsilon^{ab} \bar{\pi}_a \wedge \bar{\pi}_b -S_{0}[\bar{e}]}
\label{eqn:saddleZ}
\end{equation}
where $\bar{\pi}$ and $\bar{e}$ satisfy the saddle-point equations
\begin{equation}
    \bar{e}^{a}_{\mu}-\lambda \varepsilon^{ab}\bar{\pi}_{b,\mu} = 0 \quad   \bar{\pi}_{a,\mu} + \varepsilon_{\mu \nu} \det(\bar{e}) \langle T^{\nu}_{a} \rangle_{0} = 0 
    \end{equation}
where, in this limit, 
\begin{equation}
    \langle T^{\nu}_{a} \rangle_{0} = - \det(\bar{e})^{-1}  \frac{\delta S_{0}[e]}{\delta e^a_{\mu}} |_{e=\bar{e}}
\end{equation}
becomes the on-shell stress tensor of the undeformed theory.


As for the asymptotic boundary conditions, we echo the insight of \cite{Guica:2019nzm}. They argued that $T\bar{T}$ should be treated like any other double trace deformation in holography which lead to a change of boundary conditions at infinity (see \cite{Witten:2001ua}). In particular, the fixed dyad boundary conditions at infinity in the undeformed setting should turn into mixed boundary conditions that involve both the dyad and its radial derivative or its conjugate momentum when the deformation is turned on.

How this works is that at large $c$ (or $N$ in dimensions greater than two) the action of the deformed theory reads
\begin{equation}
S_{def}=S_{o}+\lambda \int \left(T\bar{T}\right).
\end{equation}
We then take a variation of this action to find
\begin{equation}
\delta S_{def}=\int  T^{\mu}_{(o)a}\delta e^{a}_{\mu}+\lambda \delta \int \left(T\bar{T}\right)=\int  T^{\mu}_{(\lambda)a}  \delta \tilde{e}^{a}_{\mu},
\end{equation}
where, $T^{\mu}_{(o)a}$ is the stress tensor of the seed theory which couples to its source, the dyad $e^{a}$. Similarly, $T^{\mu}_{(\lambda)a}$ is the deformed theory's stress tensor, which couples to a new source $\tilde{e}^{a}$. As was shown in \cite{Guica:2019nzm}, the latter is given by 
\begin{equation}
\tilde{e}^{a}_{\mu}=e^{a}_{\mu}-\lambda \epsilon^{ab}\epsilon_{\mu\nu}T^{\nu}_{(\lambda)b}.\label{eqn:bmix}
\end{equation}

Note that by having to hold $\tilde{e}^{a}$ fixed, the above variation vanishes. In the bulk, this is equivalent to the statement that the bulk action that $S_{def}$ is on shell and, through the holographic dictionary, also a function of solely the boundary data. In particular, the variation of the action on shell with an appropriate boundary term corresponding to the $T\bar{T}$ deformation added is given by the symplectic potential:
\begin{equation}
\delta S_{o.s.}\propto \int \pi^{\mu}_{a} \wedge \delta \tilde{e}^{a}_{\mu}, 
\end{equation}
where the RHS is integrated over the boundary, and $\pi^{\mu}_{a}$ is the momentum conjugate to the dyad $\tilde{e}^{a}_{\mu}$ induced on the boundary. The canonical transformation needed to get from the phase space parameterized  by $(e^{a},\pi_{a})$ to the $(\tilde{e}^{a},\pi_{a})$ is given by:
\begin{equation}
\tilde{e}^{a}=e^{a}-\frac{\delta W(\pi)}{\delta \pi_{a}},
\end{equation}
where 
\begin{equation}
W(\pi)=2\lambda \int \epsilon^{ab}\pi_{a}\wedge \pi_{b}. 
\end{equation}
This is indeed the boundary term in the three dimensional gravity theory that corresponds to the $T\bar{T}$ deformation. 

The specification of fixed $\tilde{e}^{a}$ boundary conditions therefore corresponds to finding some subspace of phase space on which the symplectic form computed from this potential vanishes, i.e. to a Lagrangian submanifold.

Translating the condition \eqref{eqn:bmix} into bulk language, we see that the mixed boundary condition that the $T\bar{T}$ deformation leads to is one where 
\begin{equation}
\tilde{e}^{a}_{\mu}=e^{a}_{\mu}-\lambda \epsilon^{ab}\epsilon_{\mu\nu}\pi^{\nu}_{b},
\end{equation}
is fixed at the boundary. Again, we see that the radial momentum is playing the role of the stress tensor. The function $F$ in the path integral \eqref{eqn:3dint} is therefore $\tilde{e}$ written as a function of radial derivatives of the dyad instead of the momenta. 

As mentioned before, since the state $\ket{\Psi_{\lambda}}$ satisfies the Wheeler de Witt equation, it can be computed from a radial slice arbitrarily close to the $r=r_{c}$ surface. Thus, schematically, we compute the path integral between these two surfaces as: 
\begin{equation*}
\int^{E^{(o)}|_{\infty}=(e-\epsilon * \pi)}_{E|_{r_{o}}=f} DE\, \exp{(-S_{3d}(E))}=\int_{bc} D\pi De \,\end{equation*}
\begin{equation}\exp{\left(\int_{\Sigma_{r=r_{c}}} F_{r_{o}}-\int _{\Sigma_{r=\infty}}F_{\infty}\right)}\exp\left(\int_{AdS_{3}}\pi_{a}\wedge\dot{e}^{a}\right).
\end{equation}

Here $bc$ stands for the boundary conditions at the $r=r_{c}$ surface and the surface at infinity brought to its vicinity. The functions $F_{r_{c}}$ and $F_{\infty}$ are the boundary terms at the $r=r_{c}$ surface and infinity respectively.
In first order variables, the first term is zero, and the second term is a combination of the CFT generating functional, and the term $W(\pi^{(\infty)})$ that generates the canonical transformation corresponding to the $T\bar{T}$ deformation discussed above. 
The vanishing of the Hamiltonian means that the phase space action involves only the kinetic term $\int \pi_{a}\wedge \dot{e}^{a}$. 

Then, noting that the two surfaces are arbitrarily close to each other, we can decompose the path integral over the fields $(e_{a}(r,x),\pi^{a}(r,x))$ into $((e^{(r_{c})a}(x), e^{(\infty)a}(x)), (\pi^{(r_{c})}_{a}(x), \pi^{(\infty)}_{a}(x)))$:

\begin{equation*}
=\int D\pi^{(r_{c})}De^{(r_{c})}D\pi^{(\infty)}De^{(\infty)}\exp{\left(-\int \pi^{(r_{c})}_{a}\wedge(e^{(r_{c})}-f)^{a}\right)}\times \end{equation*}
\begin{equation}
\exp{\left(-\int \pi^{(\infty)}_{a}\wedge\left(e^{(\infty)}-e^{(r_{c})}\right)^{a}\right)}\exp{\left(-\frac{\lambda}{2} \int \epsilon^{ab}\pi^{(\infty)}_{a}\wedge\pi^{(\infty)}_{b}-W^{CFT}[e^{(\infty)}]\right)}.
\end{equation}

The integral over the fields at $r=r_{c}$ can be performed straightforwardly to obtain 
\begin{equation*}
\int D\pi De\, \exp{\left(-\int \pi_{a}(e-f)^{a}-\frac{\lambda}{2}\int \epsilon^{ab} \pi_{a}\wedge \pi_{b}\right)}Z_{CFT}[e]=
\end{equation*}
\begin{equation}
\int De\, e^{- \frac{1}{2\lambda} \int \varepsilon_{ab} (f-e)^{a} \wedge (f-e)^{b}} Z_{0} [e]=Z[f].
\end{equation}
where we dropped the $(\infty)$ superscript for brevity.

\mn{E: even if we do a one step path integral here, there is still a the issue that $\int \pi \wedge \dot{e}$ }

We therefore recover the Freidel kernel formula for the $T\bar{T}$ deformed partition function.

\subsection{Differences between a 2d Partition Function and a 3d Wavefunction}
Even though the 2d partition function is in fact a 3d wavefunction, these two interpretations can lead to different physical questions.
The most important one, in our view, is the fact that a 2d expectation value is in 3d \emph{not} an expectation value but the action of an operator on a wavefunction. Consider the example of the deformed theory's stress tensor:
\begin{equation}
    \langle T^\mu_a \rangle_\lambda = -\frac{1}{\det f} \frac{\delta}{\delta f_\mu^a} \Psi_{WdW} [f].
    \label{eqn:exp-val-as-op}
\end{equation}
This leads to important differences in the set of physical operators one is allowed to consider in the two setups.

Let us recall a few basic facts about physical wavefunctions in a gauge theory.
The Hilbert space of physical wavefunctions is usually embedded in a larger Hilbert space. We may define the physical Hilbert space as the subspace of states which satisfy a set of constraints $\hat{G}$ (which generate gauge transformations),
\begin{equation}
    \mathcal{H}_{phys} = \left\{ \ket{\psi} \in \mathcal{H}_{ext}  \big{|} \ \  \hat{G} \ket{\psi} = 0 \right\}.
    \label{eqn:phys-H-defn}
\end{equation}
Consider, now, some operator $O$ which fails to commute with the constraints $\hat{G}$. This implies it does not preserve the physical Hilbert space,
\begin{equation}
    [O,\hat{G}] \neq 0 \quad \Leftrightarrow \quad O \mathcal{H}_{phys} \not\subset \mathcal{H}_{phys}.
    \label{eqn:n-g-i-op-defn}
\end{equation}
We note here, that as long as the larger Hilbert space is given, the action of the operator $O$ still gives us a state in that larger Hilbert space.
Where the gauge-invariance of the theory comes in is that, the \emph{matrix element} of this operator $O$ between two physical states is the same as the matrix element of the projection of the operator down to the gauge invariant subspace,
\begin{equation}
    \ket{\psi},\ket{\phi} \in \mathcal{H}_{phys} \quad \Rightarrow \quad \mel{\psi}{O}{\phi} = \int d\alpha \mel{\psi}{e^{-i \alpha \hat{G}} O e^{i \alpha \hat{G}}}{\phi}.
    \label{eqn:elitzur}
\end{equation}
This is the familiar statement of Elitzur's theorem.

In the 3d interpretation, the larger Hilbert space is the space of functions of vielbeins. The subspace satisfying the constraints in \eqref{eqn:constraints} defines the physical subspace.
Of particular interest to us is the momentum constraint $P^a$, which encodes 2d diff invariance.
A local operator, such as $\delta_{f_\mu^a(x)}$, does \emph{not} commute with the momentum constraints.  Therefore, in the 3d picture, it is not a physical operator.
If we took its matrix elements between two physical states as in \eqref{eqn:elitzur}, we would find that this expectation value will satisfy 2d diff-invariance. However, its action alone takes us out of the physical subspace.
The reason for stressing this is that the 2d expectation value is, in 3d language, just the action of the operator on the wavefunction as in \eqref{eqn:exp-val-as-op}.
Focusing again on 2d diff invariance, the fact that the action of the non-diff-invariant operator $O$ doesn't preserve the physical Hilbert space is just the fact that the stress tensor is only conserved up to contact terms,
\begin{equation}
    P_a O \ket{\psi} \neq 0 \quad \Leftrightarrow \quad \left\langle \grad_\mu T^\mu_a(x) O(y) \right\rangle \sim \delta(x-y).
    \label{eqn:n-g-inv-op-interpetation}
\end{equation}
The RHS is merely the statement that diff invariance is a global symmetry of the theory, and physical operators can be charged under global symmetries. 2d diff invariance being a gauge invariance of the 3d theory but a global symmetry of the 2d theory is thus perfectly consistent.

\section{$S^2$ Partition Function via a Gauge-Fixing} \label{sec:cmc}
In this section, we compute --- up to an ignorance parameter --- the $S^2$ partition function of a closely related partition function and conjecture that it is in fact the $T \bar{T}$-deformed partition function given by the Freidel kernel.
This closely related object is a particular gauge-fixing of the 3d gravity wavefunction, known as the constant mean curvature gauge.
While this partition function does solve the global $T \bar{T}$ flow equation, it does not necessarily satisfy the correct initial condition of limiting to the seed CFT partition function at $\lambda \to 0$; however, it does do so for the sphere.

To motivate this section, we point out the odd fact that the kernel satisfies the WdW equation at every point, whereas only the integrated version is related to the flow equation.
This interesting consequence is that we can write down a theory different from \eqref{eqn:kernel-intro} which satisfies the flow equation (though as we will see, not with the same initial condition).
In particular, if there exists a differential operator $D$ of the vielbein $f$  (which is the gauge fixing condition) such that
\begin{equation}
    \left[ D(\lambda,f, \delta_f), \partial_\lambda - \frac{1}{2} \int :\varepsilon_{\mu\nu} \varepsilon^{ab} \delta_{f_\mu^a} \delta_{f_\nu^b}: \right] = 0,
    \label{eqn:g-fix-fn-comm}
\end{equation}
then the partition function
\begin{equation}
    Z_{\lambda,D} [f] = P_{D} Z_\lambda [f]
    \label{eqn:gf-kernel}
\end{equation}
where $P_{D}$ is a projector onto the kernel of D, also satisfies the flow equation.
More generally, if the commutator is non-zero, we must also impose a $\delta$ function for the commutator; and then if
\begin{equation}
    \left[ \left[ D[\lambda,f, \delta_f], \partial_\lambda - \frac{1}{2} \int :\varepsilon_{\mu\nu} \varepsilon^{ab} \delta_{f_\mu^a} \delta_{f_\nu^b}: \right] , \partial_\lambda - \frac{1}{2} \int :\varepsilon_{\mu\nu} \varepsilon^{ab} \delta_{f_\mu^a} \delta_{f_\nu^b}:\right] \neq 0
    \label{eqn:d-fix-dbl-comm}
\end{equation}
we must add in a projector for that term, and so on (assuming there is enough gauge freedom to do so).
In the 3d story, this is merely the imposition of a gauge constraint.

Despite the natural 3d interpretation of this condition as a gauge-fixing of the wavefunction, it is not in general a suitable candidate for a $T \bar{T}$-deformed partition function.
The reason for this is that, even though \eqref{eqn:gf-kernel} satisfies the flow equation, its $\lambda = 0$ limit is
\begin{equation}
    Z_{\lambda,D} [f] \xrightarrow{\lambda = 0} P_{D} Z_{0} [f].
    \label{eqn:dandard-observation}
\end{equation}
This is the seed partition function if and only if the seed partition function already satsifes the condition $D$.

Of particular interest to us will be the differential operator
\begin{equation}
    D_{CMC} = f_\mu^a \delta_{f_\mu^a} - \frac{\det f}{A[f]} \int f_{\mu}^a \delta_{f_\mu^a},
    \label{cmc-diff-op}
\end{equation}
This condition enforces that the trace of the stress tensor is constant.

A more three-dimensional understanding of this condition is as follows. The gauge invariance that the Hamiltonian constraint encodes is that of local re-foliation invariance. We can fix it by choosing a specific slicing of the bulk space time into constant radius hyper-surfaces. The choice we will make is to consider radial slices of constant mean curvature. Classically, this condition states that 
\begin{equation}
K = const. 
\end{equation}
where $K=K^{\mu\nu}g_{\mu\nu}$ is the trace of the extrinsic curvature tensor. In terms of the phase space variables we are using, this condition can be re-cast as 
\begin{equation}
\varepsilon^{\mu\nu}\pi_{\mu a}f^{a}_{\nu}(x)-\frac{\det f(x)}{A[f]}\int \pi_{a} \wedge f^{a}=0. 
\end{equation}
In the quantum theory, this condition can be written as 
\begin{equation}
\left(f^{a}_{\mu}\delta_{f^{a}_{\mu}}-\frac{\det f(x)}{A[f]}\int f^{a}_{\mu}\delta_{f^{a}_{\mu}}\right)Z_{\lambda}[f]=0.
\end{equation}
Therefore this gauge fixing picks out the deformed partition function $P_{D_{CMC}}Z_{\lambda}[f]$.

The crucial point is that, for $S^2$, this gauge is special.
By symmetry, we expect the CMC condition to be true for a general diff-invariant partition function on $S^2$. In other words, by spherical symmetry, we expect the trace of the stress tensor to be constant for both the seed and the deformed theory's partition function. 
Denoting by $\bar{f}^a_{\mu}$ a family of vielbeins on an $S^2$, we can write this assumption as $P_{D_{CMC}} Z_{\lambda}[f] |_{f=\bar{f}}=Z_{\lambda}[\bar{f}] $. \footnote{The fact these are not globally, but only patch-wise defined, will not matter here. }

The local Lorentz invariance and the diffeomorphism invariance of the theory imply that the deformed partition function depends solely on the Weyl factor $\Omega$. 
In this decomposition, the CMC condition reads:
\begin{equation}
\left(\frac{\delta}{\delta \Omega(y)}-\frac{\det \bar{f}e^{2\Omega(y)}}{A}\int_{x} \frac{\delta}{\delta \Omega(x)}\right)Z_{\lambda}[\Omega]=0. 
\label{eqn:cmc-omega-form}
\end{equation}
This condition implies that $Z_{\lambda}[\Omega]$ depends only on the homogeneous mode of $\Omega$. 
This means that the partition function is a function only of the integral of $\Omega$, 
\begin{equation}
    Z_\lambda [\Omega] = Z_\lambda [\bar{\Omega}], \quad \bar{\Omega} \equiv \frac{\int (\det f) \Omega}{\int \det f},
    \label{eqn:cmc-zero-mode-dep}
\end{equation}
It is important to also understand the limitations of this statement. Let us first consider a general diff-invariant partition function on the sphere $Z_{gen}$.
Just by symmetries it is true that when the metric is that of an $S^2$
\begin{equation}
    \delta_{f_\mu^a} Z_{gen} [f] \big|_{f = \bar{f}} = \bar{f}^\mu_a \frac{\det \bar{f}}{A[\bar{f}]} \int \delta_\Omega Z_{gen},
    \label{eqn:s2-simplicity}
\end{equation}
which is the familiar statement that on an $S^2$ the stress tensor is just a constant times the metric.
However, the two point function is not so simple; this is because the two-point function of the stress tensor can be extracted from the one-point function on a slightly deformed manifold,
\begin{equation}
    \int \sqrt{g} \delta g^{\mu\nu} \langle T_{\mu\nu} T_{\alpha\beta} \rangle_g = \langle T_{\alpha\beta} \rangle_{\delta g},
    \label{eqn:two-pt-fn-defn}
\end{equation}
and the deformed manifold is not as symmetric as the sphere itself.
Another way to see this is that although the partition function satisfies \eqref{eqn:s2-simplicity}, the one-point function of the stress tensor itself doesn't.

Consider now the commutator of the CMC condition with the global flow equation: 
\begin{equation*}
\left[ D_{CMC},\partial_{\lambda}-\frac{1}{2}\int:\varepsilon_{\mu\nu}\varepsilon^{ab}\delta_{f^{a}_{\mu}}\delta_{f^{b}_{\nu}}: \right]= -:\varepsilon_{\mu\nu}\varepsilon^{ab}\delta_{f^{a}_{\mu}}\delta_{f^{b}_{\nu}}:
\end{equation*}
\begin{equation*}
 - \frac{2}{A[f]}\left(f^{a}_{\mu}\delta_{f^{a}_{\mu}}-\frac{\det f}{A[f]}\int f^{a}_{\mu}\delta_{f^{a}_{\mu}} \right)\int f^{a}_{\mu}\delta_{f^{a}_{\mu}} +\frac{ \det f}{A[f]}\int :\varepsilon_{\mu\nu}\varepsilon^{ab}\delta_{f^{a}_{\mu}}\delta_{f^{b}_{\nu}}:
\end{equation*}
\begin{equation}
=-:\varepsilon_{\mu\nu}\varepsilon^{ab}\delta_{f^{a}_{\mu}}\delta_{f^{b}_{\nu}}:-\frac{2}{A[f]}D_{CMC}\int f^{a}_{\mu}\delta_{f^{a}_{\mu}} +\frac{ \det f}{A[f]}\int :\varepsilon_{\mu\nu}\varepsilon^{ab}\delta_{f^{a}_{\mu}}\delta_{f^{b}_{\nu}}:
\end{equation}
We can now use the delta function we have already imposed to set the middle term to zero. 

 What remains then is
 \begin{equation}
\left[ D_{CMC}, \partial_{\lambda}-\frac{1}{2}\int:\varepsilon_{\mu\nu}\varepsilon^{ab}\delta_{f^{a}_{\mu}}\delta_{f^{b}_{\nu}}:\right]=-\left(:\varepsilon_{\mu\nu}\varepsilon^{ab}\delta_{f^{a}_{\mu}}\delta_{f^{b}_{\nu}}:-\frac{\det f}{A[f]}\int:\varepsilon_{\mu\nu}\varepsilon^{ab} \delta_{f^{a}_{\mu}}\delta_{f^{b}_{\nu}}:\right).
\label{eqn:ttbar-constant}
 \end{equation}
This vanishes if the $T\bar{T}$ operator has a constant expectation value.

The Freidel kernel also satisfies the local WdW equation \eqref{eqn:wdw}, $\hat{H}_{Wdw}(x)Z_{\lambda}[f]=0$, which relates the $T \bar{T}$ expectation value at a point to the expectation value of the trace and the curvature at the same point. Evaluating this expression for vielbeins on a sphere, the curvature and one point function of $trT$ are constant and hence the $T \bar{T}$ expectation value is as well. Hence, symmetry leads us to expect the commutator in \eqref{eqn:ttbar-constant}  to vanish on the sphere .\footnote{All the factors of $\det f$ are because we have found it convenient to write our differential equations in terms of scalar densities rather than scalars; these words are true of the scalars.} This would then imply that \footnote{To recap the logic here: Let $\mathcal{O} = \partial_{\lambda}-\frac{1}{2}\int:\varepsilon_{\mu\nu}\varepsilon^{ab}\delta_{f^{a}_{\mu}}\delta_{f^{b}_{\nu}}:$, then $\mathcal{O}P Z_{\lambda}[f]= P \mathcal{O} Z_{\lambda}[f] + [\mathcal{O},P]Z_{\lambda}[f]$ which vanishes at $f=\bar{f}$ as long as $[\mathcal{O},P]Z_{\lambda}[f] |_{f=\bar{f}}$=0. And this commutator indeed vanishes by using $H_{Wdw}Z_{\lambda}[f]|_{f=\bar{f}}=0$ and the assumption that spherical symmetry dictates $P_{D_{CMC}} Z_{\lambda}[f] |_{f=\bar{f}}=Z_{\lambda}[\bar{f}] $}

\begin{equation}
   \left(  \partial_{\lambda}-\frac{1}{2}\int:\varepsilon_{\mu\nu}\varepsilon^{ab}\delta_{f^{a}_{\mu}}\delta_{f^{b}_{\nu}}: \right) P_{D_{CMC}}Z_{\lambda}[f] \big{|}_{f=\bar{f}}=0
\end{equation}

If we imposed the CMC gauge generally $Z_{\lambda}[f] \rightarrow P_{D_{CMC}}Z_{\lambda}[f]$, then the residual gauge invariance can be stated as 

\begin{equation}
    P_{D_{CMC}} \hat{H}_{Wdw}  P_{D_{CMC}} Z_{\lambda}[f] = 0
    \label{eqn:reducedWDW}
\end{equation}

Evaluating this expression for $f=\bar{f}$, we therefore expect the sphere partition function to satisfy a reduced Wheeler de-Witt equation. 

It was found in \cite{Donnelly:2019pie} that, in the CMC gauge, the $T \bar{T}$ operator reduces to
\begin{equation}
P_{D_{CMC}} :\varepsilon_{\mu\nu}\varepsilon^{ab}\delta_{f^{a}_{\mu}}\delta_{f^{b}_{\nu}}: P_{D_{CMC}} Z_{\lambda}[f]=e^{-2\Omega}(\partial^{2}_{\Omega}+k \partial_{\Omega})Z_{\lambda}(\Omega). 
\label{eqn:ttbar-ode}
\end{equation}
Here $k$ parameterizes our ignorance about the details of the mode expansion one would use to turn the functional differential operator in the LHS to the partial differential operator in the RHS. Fixing $k$ would correspond to picking a specific regualtor for the coincident functional variations in the definition of $(T\bar{T})$. Indeed, this also satisfies the condition 
\begin{equation}
P_{D_{CMC}} \left(:\varepsilon_{\mu\nu}\varepsilon^{ab}\delta_{f^{a}_{\mu}}\delta_{f^{b}_{\nu}}:-\frac{\det f}{A[f]}\int:\varepsilon_{\mu\nu}\varepsilon^{ab}\delta_{f^{a}_{\mu}}\delta_{f^{b}_{\nu}}: \right) P_{D_{CMC}} Z_{\lambda}(\Omega)=0,
\end{equation}
which also arises from imposing the delta function which enforces $D_{CMC}Z_{\lambda}=0$. 

Thus, \emph{for the sphere}, we conjecture that the results of this section furnish the true deformed partition function.
The weakest link in this conjecture is the reduced Wheeler-de Witt equation \eqref{eqn:ttbar-ode} that relies on the CMC gauge (i.e. we have not been able to derive $ P_{D_{CMC}} \hat{H}_{Wdw}  P_{D_{CMC}}$ from first principles exactly, amongst other things) .
However, the ordinary differential equation \eqref{eqn:ttbar-ode} seems --- again, by symmetry --- that it is of the right form to be satisfied by the $S^2$ partition function.


\subsection{The reduced Wheeler de Witt equation}
We can now write down the Wheeler de Witt equation which $Z_{\lambda}(\Omega)$ satisfies (using more appropriate gravitational notation):
\begin{equation}
\frac{G_{N}^{2}}{2}e^{-2\Omega}\left(\partial^{2}_{\Omega}+k\partial_{\Omega}\right)Z_{\lambda}(\Omega)-\frac{1}{2}\left(\frac{e^{2\Omega}}{l^{2}}+1\right)Z_{\lambda}(\Omega)=0.
\end{equation}
It will be more convenient to write this equation in terms of the variable $r=e^{\Omega}$:
\begin{equation}
-\frac{G_{N}^{2}}{2}\left(\partial^{2}_{r}+\frac{k+1}{r}\partial_{r}\right)Z_{\lambda}(r)+\frac{1}{2}\left(\frac{r^{2}}{l^{2}}+1\right)Z_{\lambda}(r)=0. \label{redwdw}
\end{equation}
This is identical to the equation that was solved in \cite{Donnelly:2019pie}. Introducing $z=r^{2}/G_{N}$, and re-scaling $Z_{\lambda}$ to $g$ as:
\begin{equation}
g(z)=\frac{e^{z/2}}{2}Z_{\lambda}(\sqrt{G_{N}z}),
\end{equation}
we find that the reduced WdW equation \eqref{redwdw} becomes Kummer's equation:
\begin{equation}
z\partial^{2}_{z}g+\left(\frac{k}{2}+1-z\right)\partial_{z}g-a g(z)=0,
\end{equation}
where $a=\frac{1}{4G_{N}}+\left(\frac{k+2}{4}\right)$.
The general solution to this equation reads
\begin{equation}
g(z)=c_{1}\, _1 F_1\left(a,\frac{k}{2}+1,z\right)+ c_{2} z^{-\frac{k}{2}}\, _1 F_1\left(a-\frac{k}{2},1-\frac{k}{2},z\right).
\end{equation}
Boundary conditions must be chosen in order to fix $c_{1}$ and $c_{2}$, but having done that we will have a one parameter family of radial wave-functions. It turns out that for a special value of $k$, there is in fact a way to both indirectly obtain the solution to this equation and fix the boundary conditions. 

\subsection{The reduced Kernel}
For the purposes of this subsection, let us choose $k=-1$, and so the equation we want to solve is 
\begin{equation}
-\frac{G^{2}_{N}}{2}\partial^{2}_{r}Z_{\lambda}(r)+\frac{1}{2}\left(\frac{r^{2}}{l^{2}}+1\right)Z_{\lambda}(r)=0. \label{redwdw2} 
\end{equation}
Given that we want solutions to this equation with AdS asymptotics, we need that at large $r$, up to a counter term, $Z_{\lambda}(r)~\sim Z_{CFT}(r)$ where the CFT partition function solves the Weyl anomaly condition on $S^{2}$:
\begin{equation}
R\partial_{R}Z_{CFT}(R)=\frac{c}{3}Z_{CFT}(R). \label{s2anom}
\end{equation}
It turns out that we can in fact write the solution to the differential equation \eqref{redwdw2} as an integral transformation of the CFT partition function satisfying \eqref{s2anom}, which takes the form:
\begin{equation}
Z_{\lambda}(r)= \frac{e^{-\frac{2\pi r^{2}}{\lambda}}}{2}\int \textrm{d}R \left(\frac{R}{\epsilon}\right)^{b} e^{-\frac{4\pi}{\lambda}(R-r)^{2}}Z_{CFT}(R),\label{miniker}
\end{equation}
where $b$ parameterizes our ignorance of the measure factors that might have entered due to, for example ratios of determinants that the gauge fixing porcedure leads to. For our purposes, we will just assume that it is an arbitrary real parameter. The solution to \eqref{s2anom} is given by 
\begin{equation}
Z_{CFT}(R)=\left(\frac{R}{\epsilon}\right)^{\frac{c}{3}}, 
\end{equation}
and so in all, we have
\begin{equation}
Z_{\lambda}(r)=\frac{1}{2}e^{-\frac{r^{2}}{\lambda}}\int \textrm{d}R  e^{-\frac{2}{\lambda}(R-r)^{2}}\left(\frac{R}{\epsilon}\right)^{\frac{c}{3}+b}.
\end{equation}
We note that this integral can be thought of as the following Mellin transformation \footnote{Here we take the definition of the Mellin transformation to be $\mathcal{M}_{t}(f(t),s)=\frac{1}{2}\int^{\infty}_{0} dt t^{s-1} f(t)$}:
\begin{equation}
Z_{\lambda}(r)=\frac{e^{-\frac{2\pi r^{2}}{\lambda}}}{2}\mathcal{M}_{R}\left(e^{-\frac{4\pi}{\lambda}(R-r)^{2}},b+\frac{c}{3}+1\right).\label{Mellin}
\end{equation}
Which gives us:
\begin{equation*}
Z_{\lambda}(r)=2^{-\frac{b}{2}-\frac{c}{6}-3} e^{-\frac{2\pi r^2}{\lambda }} \left(\frac{\lambda}{2\pi}\right) ^{\frac{1}{6} (3 b+c)} \bigg(4 r \Gamma \left(\frac{1}{6} (3 b+c+6)\right) \, _1F_1\left(\frac{1}{6} (3 b+c+6);\frac{3}{2};\frac{4\pi r^2}{\lambda }\right)+ \end{equation*} 

\begin{equation}+\sqrt{2} \sqrt{\frac{\lambda}{2\pi} } \Gamma \left(\frac{1}{6} (3 b+c+3)\right) \, _1F_1\left(\frac{1}{6} (3 b+c+3);\frac{1}{2};\frac{4\pi r^2}{\lambda }\right)\bigg).\label{sol}
\end{equation}

This function solves the equation 
\begin{equation*}
-\frac{\lambda}{16\pi} \partial^{2}_{r}Z_{\lambda}(r)+\frac{\pi r^{2}}{\lambda}Z_{\lambda}(r) + \frac{1}{4}\left(1+2b+\frac{2c}{3}\right)Z_{\lambda}(r)=0,
\end{equation*}
which is identical to \eqref{redwdw2} if we make the identifications:
\begin{equation}
\frac{\lambda}{8\pi\left(1+2b+\frac{2c}{3}\right)}=G^{2}_{N},\,\, \frac{\lambda}{8\pi}\left(1+2b+\frac{2c}{3}\right)=l^{2}. 
\end{equation}

Like in the full kernel, if we demand that at the reduced level \begin{equation}
\lim_{\lambda\rightarrow 0}\left(e^{-\frac{r^{2}}{4G_{N}l}}Z_{\lambda}(r)\right)=Z_{CFT}(r),
\end{equation}
then we find that we have to set $b=0$. 

Now we can ask what values for $c_{1}$ and $c_{2}$ the above solution picks, and we find:
\begin{equation}
c_{1}=2^{-\frac{c}{6}-\frac{5}{2}} \left(\frac{\lambda}{2\pi}\right)^{c/6+1/2} \Gamma\left(\frac{c+3}{6}\right),\,\, c_{2}=2^{-\frac{c}{6}-\frac{5}{2}} \left(\frac{\lambda}{2\pi}\right)^{c/6} 2\sqrt{2} \Gamma\left(\frac{c}{6}+1\right).
\end{equation}
If we were using the gauge fixed three dimensional gravity path integral to generate solutions of the WdW equation \eqref{redwdw2}, then we would run into the conformal mode problem, and we would have to analytically continue in $r$ to obtain a finite result. There, we would have to choose a good contour for the integration over $r$ which is now analytically continued to complex values, and there is a one to one correspondence between the choice of this integration contour, and the values of $c_1$ and $c_2$ in the solution above. Such analytic continuation was also necessary for the analysis performed in \cite{Donnelly:2019pie} where a proper length deparameterization of the equation \eqref{redwdw2} was introduced, so that the expectation value of the length operator could be studied. Thanks to our method here, which just requires us to do a Mellin transform \eqref{Mellin}, we see that the boundary condition $\lim_{r\rightarrow \infty} e^{-r^{2}/4 G_{N}l}Z_{\lambda}(r)=Z_{CFT}(r)$ is built in, and that despite the fact that there is a path integral \ref{miniker}, it can be done without needing to analytically continue. It would be interesting to study the length operator in the setting of this kernel as well. 

To conclude this section, we reiterate our conjecture that this is the answer for the $T \bar{T}$ deformation given by the Freidel kernel, though we are far from deriving it.

\section{Towards the $S^2$ Partition Function without Gauge-Fixing} \label{sec:s2}
Having outlined the main story, we perform some explicit calculations on $S^2$ in the large-$c$ classical limit.
In the classical limit, we calculate the $S^2$ partition function, reproducing results previously found in \cite{Donnelly:2018bef}. 
Then, we make some observations about the loop expansion about this saddle-point, showing in particular that the expansion is in terms of renormalisable coupling $1/\lambda$.

\subsection{Classical Calculation} \label{ssec:cl-lim}
We will work with the effective classical action for the BS vielbein,
\begin{equation}
    S_\lambda [f,e] = \frac{1}{2\lambda} \int \varepsilon_{ab} (f-e)^a \wedge (f-e)^b - \log Z_0 [e].
    \label{eqn:eff-s-cl}
\end{equation}
Since the seed part of the action is proprtional to $c$, we find that in the limit
\begin{equation}
    c \to \infty, \quad \hat{\lambda} \equiv c \lambda < \infty
    \label{eqn:cl-limit}
\end{equation}
The action becomes
\begin{equation}
    S_\lambda = c \left\{ \frac{1}{2\hat{\lambda}} \int \varepsilon_{ab} (f-e)^a \wedge (f-e)^b - \frac{\log Z_0 [e]}{c} \right\}
    \label{eqn:S-cl-c-outside}
\end{equation}
and $c$ controls a saddle-point expansion of the path integral.

In this limit, the saddle-point equation obtained from the variation of $e$ is
\begin{equation}
    e_\mu^a = f_\mu^a - \lambda (\det e) \varepsilon_{\mu\nu} \varepsilon^{ab} \left\langle T^\nu_b [e] \right\rangle_0,
    \label{eqn:saddle-pt-eqn}
\end{equation}
where the last term is the seed stress tensor evaluated on the base space.
This equation can be pretty hard to solve in general.

However, for the case when the TS is an $S^2$, the solution must by symmetry also be an $S^2$.
With this assumption, we have
\begin{equation}
    e_\mu^a = \frac{r_{BS}}{r} f_\mu^a, \quad \log Z_0 = \frac{c}{3} \log \frac{r_{BS}}{\delta}.
    \label{eqn:s2-assumptions}
\end{equation}
The saddle-point, then, is
\begin{equation}
    r_{BS} = \frac{r}{2} + \sqrt{\frac{r^{2}}{4} + \frac{\lambda c}{24\pi}},
    \label{eqn:bs-s2-radius}
\end{equation}
and the $S^2$ partition function is
\begin{equation}
    \log Z_{\lambda} = \frac{4\pi}{\lambda} \left(   \sqrt{\frac{r^2}{4}  + \frac{\lambda c}{24\pi}} - \frac{r}{2} \right) + \frac{c}{3} \log \left( \frac{r}{2 \delta} + \frac{1}{\delta} \sqrt{\frac{r^2}{4} + \frac{\lambda c}{24\pi}} \right).
  \label{eqn:s2-partn-fn}
\end{equation}
For $\lambda >0$, this agrees with known results in the literature \cite{Donnelly:2018bef}.

Interestingly, this result is always real for the holographic sign, but becomes complex for the non-holographic sign for
\begin{equation}
    \lambda < -\frac{6\pi r^2}{c}.
    \label{eqn:s2-omplexification}
\end{equation}

\subsection{Loops} \label{ssec:one-loop}
Having performed the classical calculation, one would like to go to one-loop order around this saddle-point.
This involves expanding the fields to second order around the saddle-point and performing the Gaussian integrals.
Unfortunately, it turns out that even at this level, the calculation is fairly complicated.
However, some observations can still be made.

The simplest object to work with is not the basic form \eqref{eqn:laurent-kernel} but one in which we have integrated over target space diffeomorphisms,
\begin{equation}
    Z_\lambda [f] = \int \frac{D\xi De}{\text{vol(diff)}} e^{- \frac{1}{2\lambda} \int (f^\xi -e)^a \wedge (f^\xi - e)^b} Z_0 [e].
    \label{eqn:integrated-laurent-kernel}
\end{equation}
This is equivalent to the original kernel because of the conservation of the deformed stress tensor.
The advantage of this form is that it has a gauge symmetry given by
\begin{equation}
    \delta e^a = \mathcal{L}_y e^a, \quad \delta \xi^\mu = y^\mu
    \label{eqn:int-kernel-gauge}
\end{equation}

We will use this to fix conformal gauge
\begin{equation}
    e^a(x) = e^\Omega(x) e^{\phi(x) \varepsilon^a_{\ b}} \bar{e}^a(x), \quad \bar{e}^a \equiv \frac{r_{BS}}{r} f^a.
    \label{eqn:conf-gauge}
\end{equation}
On genus $0$ manifolds, any vielbein can be brought to this form --- there are no moduli.
There is, however, residual gauge-invariance: diffeomorphisms by conformal Killing vectors (CKVs) change the vielbein $e$ but preserve the conformal gauge \eqref{eqn:conf-gauge}.
The reason for this is that any variation with respect to a diffeomorphism can be written as
\begin{equation}
    \delta e^a_\mu = \left( \frac{1}{2} \grad_\nu \xi^\nu + \frac{1}{2} \epsilon^{\nu\rho} \grad_\nu \xi_\rho \right) e_\mu^a + (P_{1}\xi)_{\mu\nu} e^{\nu a},
    \label{eqn:e-diff-change}
\end{equation}
where $P_{1}$ is the projector onto the traceless symmetric part of the covariant derivative \cite{dhokerphong}.
The change in the vielbein due to any vector annihilated by $P_1$ is identical to that due to a local scaling and rotation.
To fix the residual gauge-invariance, we will have to restrict the $\Omega,\phi$ integrals to not vary in these directions \cite{mottola1995functional}.

First, we need the saddle-point for this new kernel.
By the same symmetry argument, the $S^2$ saddle-point is
\begin{equation}
    e^a = \frac{r_{BS}}{r} f^a, \quad \xi = 0.
    \label{eqn:s2-saddle-2}
\end{equation}
Here, $r_{BS}$ is given by \eqref{eqn:bs-s2-radius}, except with $c \to c - 24$ to account for the measure factors.

The action, in the gauge \eqref{eqn:conf-gauge}, is
\begin{align}
    S_\lambda &= \frac{1}{2\lambda} \int \varepsilon_{ab} \varepsilon^{\mu\nu} \left( f^a (x+\xi) + \partial \xi \cdot f^a (x+\xi) - e^\Omega e^{\phi \varepsilon^a_{\ c}} \bar{e}^c \right)_\mu \left( f^b (x+\xi) + \partial \xi \cdot f^b (x+\xi) - e^\Omega e^{\phi \varepsilon^b_{\ d}} \bar{e}^d \right)_\nu \nonumber\\
    &\quad - \frac{c-24}{24\pi} \int \det \bar{e} \left\{ (\partial \Omega)^2 + R[\bar{e}] \Omega \right\} - \frac{c}{3} \log \frac{r_{BS}}{\delta}.
    \label{eqn:full-action}
\end{align}
In the first line, we decided to write a general coordinate transformation as $\tilde{x}^{\mu} (x) \equiv x^{\mu} + \xi^{\mu}(x)$ (i.e. $\xi$ is not small). The second line is the CFT partition function, which can be written as the sphere partition function  with a UV regulator $ \delta $ and a Liouville action.
The shift in the central charge comes, as usual, from the measure.

There are two crucial observations that can be made from this form of the action:
\begin{enumerate}
    \item While the seed as usual gives a conformal mode problem, the kernel for positive $\lambda$ appears to soften it, and may even eliminate it entirely.
    \item The non-quadratic terms in all the fields, i.e. the interactions in an order-by-order expansion, all have coupling constant $1/\lambda$, which has mass dimension $+2$.
    This means that this gravitational theory is most likely super-renormalisable, and it should be possible to calculate the partition function with a finite number of counterterms, as in the flat space case \cite{Dubovsky:2018bmo,Aguilera-Damia:2019tpe}.
\end{enumerate}

\section{Should the Curved Space Deformation Exist?} \label{sec:arg}
In this section, we take a step back and turn our attention to the question whether we really have the right to expect the curved space $T \bar{T}$-deformed theory on arbitrary manifolds to exist at all?

We first outline several reasons why one might think it should not. 


We then present our partial case that it does, trying to be as clear as possible as to outstanding issues and highlighting what we do not understand.
Finally, these arguments motivate us to speculate whether deformations corresponding to other gravitational theories exist as well.

Broadly speaking, the two main arguments against the existence of a curved space $T\bar{T}$ deformation go as follows:
\begin{enumerate}
    \item If the deformed theory exists on arbitrary manifolds, the theory has a local stress tensor; however, we know from the Hagedorn spectrum (for at least one sign of the deformation) in flat space that the theory is not a local one, and so there should be no local stress tensor.
    \item The spectrum and S-matrix can be reproduced from a string theory description \cite{Dubovsky:2017cnj,Callebaut:2019omt}. We know that string theory places tight constraints on the allowed target space metrics. Therefore, one should not be able to define the deformed theory on arbitrary geometries.
    \item The $T \bar{T}$ operator on curved manifolds will in general have non-universal curvature-dependent contact terms, and the need to subtract these from the deforming operator precludes the ability to define the deformation in a universal manner, or even at all.
\end{enumerate}

We first start off by addressing the last point about the well-definedness of $Z_{\lambda}[f]$. 
\begin{enumerate}
    \item The WdW equation relates the local expectation value of the $T \bar{T}$ operator to that of the trace of the stress tensor.
    This means that whatever regularisation of the partition function is required to give a finite $\langle T^\mu_\mu \rangle$ also gives a finite $\langle T \bar{T} \rangle$, with the operator being defined in the simplest way as two coincident variations.
    
    We have not been able to give a fully satisfactory argument, however, that the partition function can be regularised to give a finite $\langle T^\mu_\mu \rangle$ with a finite number of counterterms.
    
    \item A partial argument, however, is the order-by-order expansion in section \ref{ssec:one-loop}.
    We noted there that the expansion seems to be controlled by a super-renormalisable coupling $\lambda^{-1}$, and therefore it should be possible to regularise the partition function in a suitable manner.
    
    \item Another partial argument is our conjecture that the results of section \ref{sec:cmc} in fact furnishes the full finite $c$ partition function on an $S^2$.
    
\end{enumerate}

Having presented our evidence that the partition function is sensible, it is worth answering the other, somewhat less specific objections. 

About the connection to string theory, we note that the constraints on the target space of string theory come from a Weyl gauge symmetry; we have not found anything to take its place here.
If the deformed theory is exactlty string theory, it is reasonable to claim that one can't wiggle the target space; and given that the two theories have the same spectrum and S-matrix, it seems reasonable to say that they are the same.
However, one of the properties that defines a theory is the set of legal deformations; so, unless there is an inconsistency in the Freidel path integral, the two theories do differ in this manner!
Of course, the question is whether there are any inconsistencies, and the answer is none that we know of.

As to the first objection, we first point out that the theory possessing a local stress tensor is not inconsistent with it being non-local.
Here, it is useful to make a distinction between two types of non-locality.
The type of non-locality in standard gravity theories means that there are no local operators, in the sense that there are no physical operators that can be assigned to points. This underlies the intuition that non-locality of the theory precludes the existence a local stress tensor. The fact that the target space metric $f$ on which the deformed theory lives is, as far as we know, non-dynamical suggests that this is most likely not the sort of non-locality relevant for $T\bar{T}$. 

Non-locality can also be purely \emph{algebraic}, in the sense that one \emph{can} assign operators to points, but their algebraic relations with other operators encode the non-locality. Odd parity fermionic operators furnish a very simple example of this. Two such operators fail to commute when they are space-like separated (they anticommute instead). 
This would mean that the non-locality of the theory is manifested not in the lack of a local stress tensor, but in its correlation functions.
We note that this sort of non-locality has already been seen in the contexts of direct study of the $T\bar{T}$ deformation \cite{Cardy:2019qao}, holography \cite{Lewkowycz:2019xse} and algebraic QFT \cite{Lechner:2006kb} contexts.

In conclusion, while our rebuttal is partial at best and these important objections certainly require further study, we do have reason for hope. 

\section{Conclusion \& Future Directions} \label{sec:conc}

In this paper, we have proposed a generalization of the $T \bar{T}$ deformation to curved spaces. The original definition of the deformation in flat space rested upon the elegant results of \cite{Zamolodchikov:2004ce, Smirnov:2016lqw} on the well-definedness of the operator $T \bar{T}$. This operator perspective sets a long list of expectations. Past work, such as  \cite{Jiang:2019tcq}, showed the obvious hurdles to satisfying these expectations in the curved space setting.  

Our approach to the $T \bar{T}$ deformation centers instead around a flow equation satisfied by the partition function. This equation may be solved in terms of a path integral kernel convolved with the seed's partition function. By writing this kernel in terms of the exponential of some action, we can interpret the deformation as coupling the seed to a theory of 2d topological gravity.

Two main considerations guided our choice of an appropriate generalization of the $T\bar{T}$ flow equation. The first was the existence of the Freidel integral transform and the flow equation it could be shown to satisfy. Secondly, we required our result to reproduce the DGHC kernel in the flat space limit - which it does. In particular, based on the torus computation of \cite{Dubovsky:2018bmo}, our proposal therefore reproduces the dressing of the energy levels known from the inviscid Burger's equation.

Freidel's proof that the $Z_{\lambda}[f]$ satisfies the local Wheeler-de-Witt equation(s) of $AdS_3$ gravity has far reaching consequences for us. First, it promotes the standard large-c ``trace flow'' equation to an exact statement. Secondly, relating the local expectation value of $T \bar{T}$ \footnote{As always, by this we mean the result of acting on $Z_{\lambda}[f]$ by $\frac{1}{2}\int_{\Sigma} d^2 x \varepsilon^{ab} \varepsilon_{\mu \nu }  :\frac{\delta} {\delta f_{\mu}^{a}(x)} \frac{\delta}{ \delta f_{\nu}^{b} (x)}:$} to the one point function of the trace of the stress tensor matches the divergence structure of the two operators. This has not sufficed to demonstrate the total well-definedness of the deformation; however, $\tr  T$ is admittedly a simpler beast to tame. Thirdly, it is one way to derive the kernel's ability to capture a local flow equation with $\lambda \rightarrow \lambda(x)$. 

A particularly satisfying feature of our proposal is its ability to unify and explain several important results in the field. The composition property of the kernel immediately explains the similarity between Cardy's infinitesimal analysis and the finite $\lambda$ $JT'$ proposal of DGHC; the intuitive derivation via the Legendre transformation makes it even more obvious that there is no fundamental difference between the infinitesimal and finite deformation cases. The equivalence between the deformed theory's partition function and a radial wavefunction in $AdS_3$ justifies the appealing interpretation of the deformation of a holographic theory as flowing into the bulk, only in the stress tensor sector. It is indeed the classical limit of this quantum theory of pure general relativity. Finally, via a semiclassical treatment of the phase space path integral representation of the kernel, we landed on the boundary conditions discussed in \cite{Guica:2019nzm}. 

This paper raises as many questions as it answers. The dream would of course be to evaluate $Z_{\lambda}$ in some analytically tractable fashion. The lack of a quotient by $\text{Vol(Diff)}$ in the kernel renders explicit calculations tedious. The localization to zero modes in \cite{Dubovsky:2018bmo} traces its origins back to the simplicity of diffeomorphisms in flat space. In curved space, such a reduction appears too much to hope for. 

However, the theory still carries hints of an underlying simplicity.
The main one is the WdW equation; an equation of that form is expected to be valid for relevant perturbations, but this theory satisfies it for an irrelevant one, exhibiting that it is a very special deformation.
Understanding the implications of this equation will undoubtedly lead to much progress.

While the connection to 3d gravity is not a holographic one, one might nonetheless wonder whether the current correspondence might be extended to reproduce the 1-loop effect of matter fields. While preliminary, we suspect it can.

Consider a 3d gravity theory coupled to bulk matter.
On a radial slice, there are two objects coming from the 2d scalar, the operator $J$ and its conjugate momentum $O$.
One can then define a wavefunction that satisfies, schematically,
    \begin{equation}
        \left\{ f_\mu^a \delta_{f_\mu^a} +  \lambda \varepsilon_{\mu\nu} \varepsilon^{ab} \delta_{f_\mu^a} \delta_{f_\nu^b} + \frac{c-24}{48\pi} \det f R[f] + \hat{t}^\rho_\rho [J,\delta_J] \right\} \psi[f,J] = 0.
       \label{eqn:modified-H}
    \end{equation}
We can then interpret this wavefunction as a 2d deformed theory, with sources $f,J$ for the operators $T,O$.
    
Further, one can regard $T \bar{T} + t^\rho_\rho$ as the 2d deforming operator. We might regard this as a quantum generalization of the large $c$ proposal in \cite{Hartman:2018tkw,Taylor:2018xcy}; the full deforming operator, if not the individual pieces, may be easy to control in this theory since the UV divergences of the deforming operator are the same as that of the trace of the stress tensor. Interestingly, a similar wavefunction for higher-spin gravity has already been found in  \cite{Rashkov:2019wvw}. We hope to report on progress on these issues in the near future.

\section*{Acknowledgments}
We would like to thank Jeremias Aguilera-Damia, Louise Anderson, Steven Carlip, Evan Coleman, Daniel Freedman, Laurent Freidel, Abhijit Gadde, Victor Gorbenko, Aitor Lewkowycz, Jorrit Kruthoff, Onkar Parrikar, Daniel Ranard, Aldo Riello, Eva Silverstein, Andrew Tolley, Gonzalo Torroba and Herman Verlinde for discussions and guidance.
We would also like to thank the authors of \cite{Aguilera-Damia:2019tpe} for collaboration on related work.
Parts of this work were done during the programs ``Quantum Information and String Theory 2019'' at Kyoto University, ``Student Talks on Trending Topics in Theory (ST${}^4$) 2019'' at IISER Bhopal and the working group on $T\bar{T}$ at the Aspen Centre for Physics.
This research was supported in part by Perimeter Institute  for Theoretical  Physics. Research  at  Perimeter  Institute  is  supported  in  part  by the  Government  of  Canada  through  the  Department  of  Innovation,  Science  and  Economic Development  Canada  and  by  the  Province  of  Ontario  through  the  Ministry  of  Economic Development, Job Creation and Trade.

\begin{appendices}
  \section{Facts about Vielbeins} \label{app:e}
In this section, we provide a short introduction to vielbeins for the uninitiated.

Instead of parametrising the geometry of the manifold in terms of the metric at every point, we can parametrise it in terms of a local inertial frame -- the vielbeins.

The metric can then be written in terms of the vielbeins as
\begin{equation}
    g_{\mu\nu} = e_\mu^a e_\nu^b \eta_{ab},
    \label{eqn:e-defn}
\end{equation}
where $\eta$ is the Euclidean or Minkowski flat metric depending on signature.
All $\mu,\nu$ indices are raised and lowered by the metric $g$, and $a,b$ indices with $\eta$.

Some useful relations are
\begin{align}
    e^\mu_a e_\mu^b = \delta_a^b, \quad e^\mu_a e^a_\nu &= \delta^\mu_\nu, \quad e^\mu_a = g^{\mu\rho} \eta_{ac} e_\rho^c \nonumber\\
    \det e = \frac{1}{2} \varepsilon_{ab} \varepsilon^{\mu\nu} e_\mu^a e_\nu^b &= \sqrt{|g|} \nonumber\\
    e^1 \wedge e^2 \equiv e^1_\mu e^2_\nu dx^\mu \wedge dx^\nu &= \sqrt{|g|} d^2 x.
    \label{eqn:e-basic-props}
\end{align}

A covariant derivative can be defined for mixed-index tensors by
\begin{equation}
  \grad_{\mu} A^{\nu a}_{\ \ b} = \partial_{\nu} A^{\mu a}_{\ \ b} + \omega_{\mu\ c}^{a} A^{\nu c}_{\ \ b} - \omega_{\mu\ b}^{c} A^{\nu a}_{\ \ c} + \Gamma^{\nu}_{\ \mu\lambda} A^{\lambda a}_{\ \ b}.
  \label{eqn:spin-connection-defn}
\end{equation}
The manifold indices are transported as usual by the Christoffel symbol, and the tangent space indices are transported by the \emph{spin connection} $\omega$.
The spin connection can be defined by the \emph{torsionlessness} constraint,
\begin{equation}
    de^a + \omega^a_{\ b} \wedge e^b = \frac{1}{2} dx^1 \wedge dx^2 \varepsilon^{\mu\nu} (\partial_\mu e_\nu^a + \omega_{\mu\ b}^{\ a} e_\nu^b) = 0.
    \label{eqn:torsionlessness}
\end{equation}
Along with the usual definition of the Christoffel symbol, this means that the covariant derivative of the vielbein vanishes,
\begin{equation}
    \grad_\mu e_\nu^a = 0.
    \label{eqn:metric-compatibility}
\end{equation}
In two dimensions, we have the simplification that
\begin{equation}
    \omega_{\mu\ b}^{\ a} \equiv \omega_\mu \varepsilon^a_{\ b}.
    \label{eqn:2d-omega}
\end{equation}

This entails that the curvature two-form is given simply by $d\omega$, which is related to the Ricci scalar via 

\begin{equation}
    \sqrt{g}R=\varepsilon^{\mu \nu} \partial_{\mu } \omega_{\nu}
\end{equation}
  
  \section{Details of the $De$ measure} \label{app:De}
  
  The measure we use was suggested to us by A. Tolley. It is somewhat non-standard. It therefore deserves a more in-depth discussion than the main text could afford. The most important calculation in this section guarantees the DGHC kernel with our choice of measures reproduces the correct torus path integral.
  
  Let us first highlight the importance of a translation invariant measure for the $\lambda \rightarrow 0$ limit. In that limit, we should recover the seed: $\lim_{\lambda \rightarrow 0} Z_{\lambda}[f] = Z_{0}[f]$.

Define the rescaled and shifted vielbein $\sqrt{\lambda } h^a= (e-f)^a $. Translation invariance of the measure gives $De=Dh$. We have dropped the (infinite) constant $\det(\sqrt{\lambda})$,   which we can absorb via a local counterm, in the form of a renormalization of the cosmological constant \cite{polchinski1986evaluation}. The definition of the measure implies $\int Dh e^{- \frac{1}{2}(h,h)}= 1$. Taking the limit of the partition becomes trivial now 
\begin{equation}
    \lim_{\lambda \rightarrow 0 }Z_{\lambda}[f]= \int Dh e^{-\frac{1}{2} \int \varepsilon_{ab} h^a \wedge h^b} \lim_{\lambda \rightarrow 0} Z_{0}[\sqrt{\lambda}h+f]= Z_{0}[f]
\end{equation}

The standard, non-translation invariant measure relies on the inner product 

\begin{equation}
    (\delta e,\delta e)_{e}=\int \delta_{ab} \delta e^a \wedge \star_{e} \delta e^b = \int \delta_{ab} \delta e^a_{\mu } \delta e^b_{\nu} \left( e^{\mu}_{c} e^{\nu}_{d}\delta^{cd} \right) \det(e) d^2 x
\end{equation} 

which clearly depends on the point in field space $e$ where it is evaluated. Many of the results we used about the anomalous Weyl scaling of the measure were originally derived for this choice of measure. To make contact with this past literature, we decompose our $De$ in terms of a local Weyl mode $\Omega$, a local angle $\phi$ and a diffeomorphism generated by a vector field $\xi$. 

To understand the measure written in terms of these variables, first decompose the variation of a vielbein around a point in field space labeled by $e^{a}=\left[e^{\Omega} \left( e^\varepsilon \right)^{a}_{b} \hat{e}^{a} \right]^{\xi}$,

\begin{eqnarray*}
\delta e^{a} \large{|}_{e^{a}=\left[e^{\Omega} \left( e^\varepsilon \right)^{a}_{b} \hat{e}^{a} \right]^{\xi}} & = & \left(\delta\Omega+\frac{1}{2}\star_{e}d\star_{e}\delta\xi\right)e^{a}+\varepsilon_{b}^{a}\left(\delta\phi+i_{\delta\xi}\omega+\frac{1}{2}\star_{e}d\delta_{\xi}\right)e^{b}\\
& & + e_{\nu}^{a}\frac{1}{2}\left(\nabla_{\mu}^{(e)}\delta\xi^{\nu}+\nabla^{(e)\nu}\delta\xi_{\mu}-\delta_{\mu}^{\nu}\nabla_{\alpha}^{(e)}\delta\xi^{\alpha}\right)dx^{\mu}\\
\end{eqnarray*}

Inserting this into the $e$ inner product, we find 

\begin{align}
    (\delta e,\delta e) & = \int \varepsilon_{ab} \delta e^a \wedge \delta e^b \\
    & = \int \left(\delta\Omega+\frac{1}{2}\star_{e}d\star_{e}\delta\xi \right)^2 \varepsilon_{ab} e^a \wedge e^b + \int \left(\delta\phi+i_{\delta\xi}\omega+\frac{1}{2}\star_{e}d\delta_{\xi}\right)^2 \varepsilon_{ab} e^a \wedge e^b \\
    & + \int g^{\alpha \beta}[e] g^{\mu \nu}[e] (P_{1} \delta \xi)_{\alpha \mu } (P_{1} \delta \xi)_{\beta \nu } \varepsilon_{ab} e^a \wedge e^b 
\end{align}

\mn{was careful that $\Omega$ and $\phi$ crossterms vanish, but double check cross terms with $\delta \xi $! YEAH STUFF MISSING HERE, WILL FIX LATER}
where we defined $(P_1 \delta \xi )_{\mu \nu} = \frac{1}{2}\left(\nabla_{\mu}^{(e)}\delta\xi^{\nu}+\nabla^{(e)\nu}\delta\xi_{\mu}-\delta_{\mu}^{\nu}\nabla_{\alpha}^{(e)}\delta\xi^{\alpha}\right) $.

Defining the standard scalar measures $D\Omega $ and $D\phi $ through 

\begin{align}
    \int D\delta \Omega e^{\frac{1}{2}(\delta \Omega, \delta \Omega)}=1 \quad (\delta \Omega, \delta \Omega) = \int (\delta \Omega)^2  \varepsilon_{ab} e^a \wedge e^b\\
     \int D\delta \phi e^{\frac{1}{2}(\delta \phi, \delta \phi)}=1 \quad (\delta \phi, \delta \phi) = \int (\delta \phi)^2 \varepsilon_{ab} e^a \wedge e^b
\end{align}

alongside the standard vector measure  $D\xi$ 

\begin{align}
    \int D\delta \xi e^{\frac{1}{2}(\delta \xi, \delta \xi)}=1 \quad (\delta \xi, \delta \xi) = \frac{1}{2} \int \delta^{cd} e_c^{\mu} e_d^{\nu}(\delta \xi _{\mu}) (\delta \xi _{\nu}) \varepsilon_{ab} e^a \wedge e^b\\
\end{align}

we may calculate the Jacobian translating between $De$ and $D\Omega D\phi D\xi $ via 

\begin{align}
 \int D\delta e e^{-\frac{1}{2}(e,e)} & = J \int D\delta \Omega D\delta \phi D\delta \xi  e^{-||\delta\Omega+\frac{1}{2}\star_{e}d\star_{e}\delta\xi||^2 - ||\delta\phi+i_{\delta\xi}\omega+\frac{1}{2}\star_{e}d\delta_{\xi}||^2 - (\delta \xi, P_1^{\dagger} P_{1} \delta \xi) } \\
 & = J \int D \delta \xi  e^{-\frac{1}{2}(\delta \xi, P_1^{\dagger} P_{1} \delta \xi) } \\
 & =  J \ \text{Vol}(Ker P_1) \left( \det'(P_1^{\dagger} P_{1}) \right) ^{-1/2} \\
\end{align}

where $det'$ denotes the exclusion of zero modes. In going to the second line, we shifted the $ \Omega $ and $\phi$ integrals by $\frac{1}{2}\star_{e}d\star_{e}\delta\xi$ and $+i_{\delta\xi}\omega+\frac{1}{2}\star_{e}d\delta_{\xi}$, respectively. We thus arrive at the final relation 

\begin{equation}
 De = D \Omega D\phi D\xi \frac{\sqrt{\det'(P_{1}^{\dagger}P_{1})}}{\text{Vol}(Ker P_1)}
 \label{eqn:stringjac}
\end{equation}

\begin{equation}
  D_{e} e = D_{e} \Omega D_{e} \phi D_{e} \xi \Delta_{FP}
  \label{eqn:e-measure}
\end{equation}
We want to calculate the central charge dictating the anomalous scaling of this measure. First off, Myers-Periwal \cite{Myers:1992ea} showed that $c [D\Omega D\phi] = 2$. The Jacobian appearing in \eqref{eqn:stringjac} is well-known from string theory, and contributes  $c[\Delta_{FP}] = -26$.
To evaluate the central charge of $D\xi$, we write
    \begin{equation}
      \xi = d\alpha + *d\beta + \xi_{m}, \quad \xi_{m} \in H^{1}
      \label{eqn:xi-exp}
    \end{equation}
Thus, we find that
    \begin{equation}
      D\xi = D'\alpha D'\beta D\xi_{m} \det'(\Box).
      \label{eqn:d-xi-exp}
    \end{equation}
Hence, 
    \begin{equation}
      c[D\xi] = c[D\alpha] + c [D\beta] + c[\det'\Box] = 1 + 1 - 2 = 0.
      \label{eqn:d-xi-c}
    \end{equation}
We have ignored all zero-mode issues here, since they are irrelevant in determining the central charge.
Thus, we find that
\begin{equation}
  c[De] = 2 - 26 = -24.
  \label{eqn:De-c}
\end{equation}

  \section{A Result for the Modified WdW Eqn with $\lambda \rightarrow \lambda(x)$} \label{app:modWDW}
  
 In this section, we provide a few more steps in deriving the modified WdW equation for $\lambda \rightarrow \lambda(x)$. 
 
 We start out with \eqref{eqn:mod-SD-diffs} which we write as 
 
 \begin{equation}
     \varepsilon^{\mu \nu } \left( \partial_{\mu} f^{a}_{\nu} + \varepsilon^{a}_{b} \langle \omega_{\mu}[e]  \rangle f^{b}_{\nu} \right) = \varepsilon^{\mu \nu } \left( \langle e^{a}_{\mu}-f^{a}_{\mu }\rangle \partial_{\nu} \log(\lambda) \right)
 \end{equation}
 
 Then by adding and subtracting $\varepsilon^{\mu \nu} \varepsilon^{a}_{b} \omega_{\mu}[f] f^{b}_{\nu}$ and using the torsionlessness condition for $ \omega_{\mu}[f]$, this becomes
 
 \begin{equation}
     \varepsilon^{a}_{b}\varepsilon^{\mu \nu} \left( \langle \omega_{\mu}[e]\rangle- \omega_{\mu}[f]\right) f^{b}_{\nu}= \varepsilon^{\mu \nu } \frac{(e-f)^{a}_{\mu}}{\lambda(x)}\partial_{\nu}\lambda(x) = \varepsilon^{ac}\frac{\delta \log Z_{\lambda}}{\delta f^{c}_{\nu}}\partial_{\nu}\lambda(x)  
 \end{equation}
 
 By multiplying each side by $f_{a,\sigma}$ and noting that $\varepsilon_{ab}f^{a}_{\sigma}f^{b}_{\nu}= \det(f) \epsilon_{\sigma \nu}$, we obtain 
 
 \begin{equation}
     \langle \omega_{\sigma}[e] \rangle = \omega_{\sigma}[f] + \frac{1}{\det(f)}f_{a,\sigma} \varepsilon^{ac} \frac{\delta \log Z_{\lambda}}{\delta f^{c}_{\nu}}\partial_{\nu}\lambda(x)  
 \end{equation}
 
 Using $\varepsilon^{\alpha \sigma} \partial_{\alpha} \omega_{\sigma}[f]=\det(f)R[f]$ and similarly for $\omega[e]$, we therefore arrive at 
 
 \begin{equation}
     \langle \det(e) R[e] \rangle Z_{\lambda }[f]= \det(f) R[f] Z_{\lambda}[f] + \varepsilon^{\alpha \sigma} \partial_{\alpha} \left( \frac{1}{\det(f)}f_{a,\sigma} \varepsilon^{ac} (\partial_{\nu}\lambda) \frac{\delta}{\delta f^{c}_{\nu}(x)} \right) Z_{\lambda}[f]
 \end{equation}
 
 as quoted in the main text.

\end{appendices}

\bibliographystyle{JHEP}
\bibliography{refs}

\end{document}